\documentclass[aps,prb,twocolumn,groupedaddress]{revtex4-1}

\usepackage{mathrsfs}
\usepackage{amsmath}
\usepackage{amssymb}
\usepackage{graphicx}

\begin{document}

\title{Charge Order in NbSe$_{2}$}

\author{Felix Flicker}
  \email{flicker@physics.org}
  \affiliation{Department of Physics, University of California, Berkeley, California 94720 USA}
\author{Jasper van Wezel}
  \email{vanwezel@uva.nl}
  \affiliation{Institute for Theoretical Physics, Institute of Physics, University of Amsterdam, 1090 GL Amsterdam, The Netherlands}

\date{\today}

\begin{abstract}
We develop in detail a model of the charge order in NbSe$_2$ deriving from a strong electron-phonon coupling dependent on the ingoing and outgoing electron momenta as well as the electronic orbitals scattered between. Including both dependencies allows us to reproduce the full range of available experimental observations on this material. The stability of both experimentally-observed charge-ordered geometries (1Q and 3Q) is studied within this model as a function of temperature and uniaxial strain. It is found that a small amount of bulk strain suffices to stabilize the unidirectional order, and that in both ordering geometries, lattice fluctuations arising from the strong electron-phonon coupling act to suppress the onset temperature of charge order, giving a pseudogap regime characterized by local order and strong phase fluctuations. 
\end{abstract}

\maketitle

\section{Introduction \label{sec:CDW Introduction}}
The Peierls instability is well-known to give rise to charge density wave (CDW) order in quasi-one dimensional materials~\cite{Peierls}. In dimensions higher than one, the electronic susceptibility does not generically diverge, and the development of CDW order requires either a nested Fermi surface, or an enhancement of the momentum-dependent electron-phonon coupling. Niobium diselenide ($2H$-NbSe$_{2}$) is an early example of a quasi-two dimensional material demonstrating a charge density wave transition~\cite{WexlerWoolley76}. It is part of the extensive family of transition metal dichalcogenides, which includes many more bulk charge-ordered materials, such as TaS$_{2}$, TaSe$_{2}$, NbS$_{2}$ and TiSe$_{2}$~\cite{DoranEA78,DoranEA78b,vanWezelEA10,vanWezel11}. The charge order in these compounds, and in NbSe$_{2}$ in particular, has been suggested to serve as a model system for the charge order recently observed in cuprate high-$T_{C}$ superconductors, including Bi$_2$Sr$_2$CaCu$_2$O$_{8+x}$ (Bi2122)~\cite{HoffmanEA02,HoffmanEA02b,ParkerEA10}, Ba$_{1-x}$Na$_{x}$Ti$_{2}$Sb$_{2}$O~\cite{DoanEA12}, and YBa$_2$Cu$_3$O$_7$ (YBCO)~\cite{RossnagelEA01,ChangEA12}.

The comparison is of particular interest, since a CDW in dimensions higher than one not only breaks translational symmetry but also the rotational symmetry of the underlying lattice, selecting a preferential direction in space. Aside from forming a CDW with one fixed wave vector (`1Q CDW') it is then possible to form multiple coexisting CDWs. In the layered cuprate superconductors, with a square lattice of copper atoms, the charge order may be either a 1Q or a 2Q CDW, with the latter consisting of 1Q CDWs along both in-plane lattice directions. These two phases can even compete with one another~\cite{MelikyanNorman14}. In comparison, the well-known 3Q CDW in the layered hexagonal material NbSe$_{2}$, consisting of three superposed density waves at relative angles of $2\pi/3$, has recently been shown to compete with 1Q order in locally-strained regions on the material's surface~\cite{SoumyanarayananEA13}.

In the absence of nesting, understanding the influence of electron-phonon coupling is paramount to understanding both the origin of charge order, and the competition between different geometries of CDW states. In two recent papers, the present authors outlined a model describing the charge-ordered phases of niobium diselenide~\cite{FFJvWNatComms15,FFJvWPRB15}. The present paper extends these results and provides the calculational details of the model, showing how both the momentum and orbital dependence of the electron-phonon coupling co-operate to drive the onset of charge order in NbSe$_{2}$, and how their interplay with the electronic structure directs the competition between single-Q and multi-Q CDWs.

\subsection*{Niobium diselenide}

\begin{figure}
\begin{centering}
\includegraphics[width=\columnwidth]{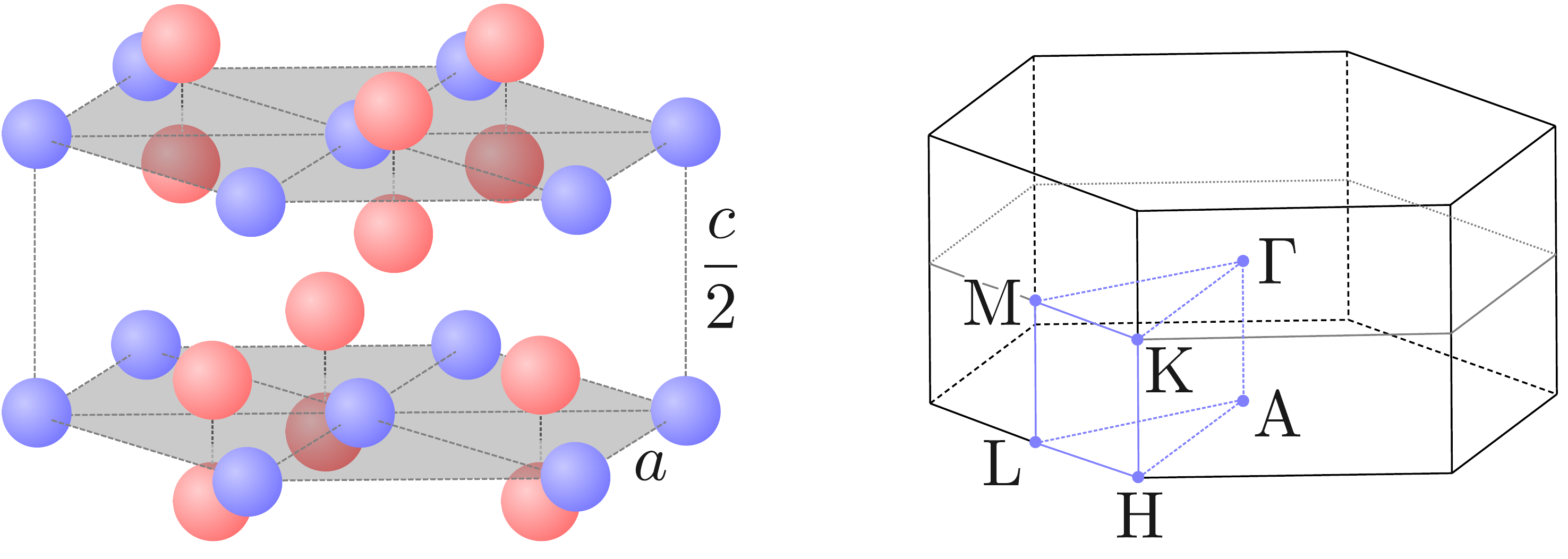}
\end{centering}
\caption{\label{fig:NbSe2 structure} {\bf Left} the crystal structure of $2H$-NbSe$_{2}$. Niobium atoms are shown in blue and selenium atoms in red. Each primitive unit cell contains two formula units, since the positions of the seleniums differ between the two pictured `sandwiches'. {\bf Right} the Brillouin zone with high-symmetry points indicated. The image is stretched disproportionately along $\Gamma$A for clarity - in fact the crystal's $c/a$ ratio is approximately $18.1/6.5$~\cite{WexlerWoolley76,DoranEA78}.}
\end{figure}

The $2H$ polytype of niobium diselenide has a layered, hexagonal crystal structure, space group $P6_{3}/mmc$ ($D_{6h}^{4}$), with two niobium atoms per primitive unit cell~\cite{WexlerWoolley76,MonctonEA75,DoranEA78}. The unit cell, shown in Fig.~\ref{fig:NbSe2 structure}, includes two `sandwiches' each consisting of two layers of selenium atoms enclosing a single layer of niobium atoms. In the lower sandwich, three of the six interstices in the hexagonal niobium layer have selenium atoms above and below them, while in the upper sandwich the other three interstices are selected. The large anisotropy, signaled by a small ratio of interlayer to intralayer coupling, suggests that quasi-two dimensional models can be expected to capture the important physics~\cite{DoranEA78,Doran78}.

NbSe$_{2}$ undergoes a phase transition to a 3Q incommensurate CDW state at $33.5\,$K. From neutron diffraction and high-resolution X-ray scattering the CDW wave vectors are known to sit along the ${\Gamma}$M directions at $\mathbf{Q}_{\text{CDW}}=\left(1-\delta\right)\frac{2}{3}\Gamma$M, with $\delta\approx0.014$~\cite{MonctonEA75,MonctonEA77,FengEA15}. The 1Q CDW phases found in locally-strained regions by recent scanning tunneling microscopy (STM) experiments have a slightly modified wave number of $\mathbf{Q}_{\text{CDW}} = \left(1-\delta\right) \frac{2}{3}\Gamma$M, with $\delta\approx0.143$~\cite{SoumyanarayananEA13}. 

\begin{figure}
\begin{centering}
\includegraphics[width=\columnwidth]{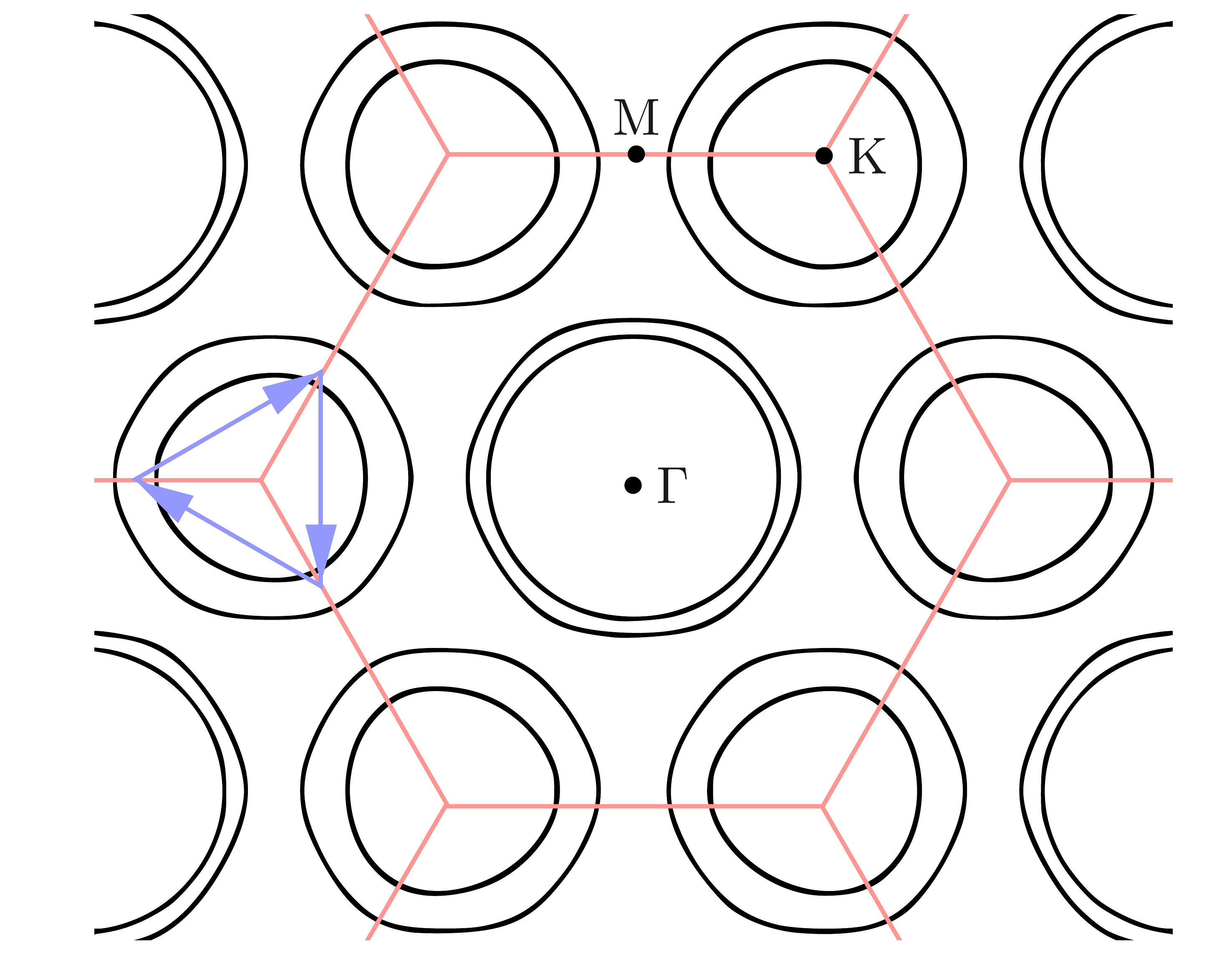}
\end{centering}
\caption{\label{fig:FS} The Fermi surface of NbSe$_{2}$ within the $\Gamma$MK plane. The Brillouin zone is marked in red, with high symmetry positions indicated. In blue are the three experimentally-observed CDW wave vectors~\cite{RahnEA12,WexlerWoolley76,RossnagelEA01}. CDW gaps have been observed to open on parts of the inner Fermi surface pockets surrounding the K-points, in the three regions approximately connected by the CDW wave vectors~\cite{BorisenkoEA09,RahnEA12}.}
\end{figure}

The Fermi surface of NbSe$_{2}$, shown in Fig.~\ref{fig:FS}, does not contain strongly-nested regions. This has lead to a number of proposals for alternative driving mechanisms underlying the CDW formation, including nested saddle-points in the electronic dispersion~\cite{RiceScott75}, local field effects~\cite{JohannesEA06}, or a combination of weak nesting with strongly momentum-dependent electron-phonon coupling~\cite{Doran78,WeberEA11}. The latter claim found some support in the observation that inelastic X-ray scattering indicates that the longitudinal acoustic phonons in NbSe$_2$ are softened over a broad range of momenta around the CDW wave vector~\cite{WeberEA11,WeberEA13}. This is in contrast to the sharp Kohn anomaly characteristic of one-dimensional and well-nested CDW materials.

A CDW gap is seen to open in Angle Resolved Photo-Emission Spectroscopy (ARPES) experiments, but only at select points on the Fermi surface, located on the inner pockets surrounding the K-point. The material therefore remains metallic below the CDW transition, and displays Fermi arcs reminiscent of those seen in high-$T_{C}$ superconductors~\cite{BorisenkoEA09,RahnEA12,RossnagelEA01,KissEA07}. As in the cuprates, the presence of Fermi arcs signals a pseudogap phase, in which the density of states is reduced for a sizable span of temperatures above $T_{\text{CDW}}$, while strong local fluctuations of the atomic positions can be observed within the same temperature range~\cite{ChatterjeeEA14}.

In addition to the nature of the pseudogap phase, puzzling aspects of the NbSe$_{2}$ phenomenology include the question of why only one band develops a CDW gap; uncertainty over the size of the CDW gap, with estimates ranging from $0\,\mbox{meV}$~\cite{BorisenkoEA09} to $35\,\mbox{meV}$~\cite{HessEA91}; and an observed asymmetry in the particle and hole states near the CDW gap~\cite{SoumyanarayananEA13}, leading to an offset of the gap with respect to $E_{\text{F}}$~\cite{ArguelloEA14}. The present authors recently outlined a model based on a strongly momentum- and orbital-dependent electron-phonon coupling to explain all of these unusual features of the NbSe$_2$ CDW state, as well as providing a quantitative explanation of how the wave vectors of the 3Q and 1Q CDW instabilities are selected~\cite{FFJvWNatComms15,FFJvWPRB15}. 

The present paper extends the results of this previous work as well as providing details of the calculations involved. In Section \ref{sec:NbSe2 band structure} we carry out a tight-binding fit to establish the NbSe$_{2}$ band structure and its orbital make-up, from which we derive an analytic form for the electron-phonon coupling in Section \ref{sec:Varma}. In Section \ref{sec:nesting} we combine these ingredients into a microscopic model which we use to quantify the extent of Fermi surface nesting, and its contribution to driving the CDW transition relative to the momentum dependence of the electron-phonon coupling. In Section \ref{sec:The-CDW-Gap} we compare the predictions of the microscopic model regarding the gap shape and size to known experimental results, and in Section \ref{sec:Higher-Order-Diagrams} we present the mean-field phase diagram as a function of both temperature and uniaxial strain. Finally, in Section \ref{sec:pseudogap} we include fluctuations beyond the mean field using the Mode-Mode coupling Approximation (MMA), and use this to describe the pseudogap regime which emerges above the charge ordering transition. We consider the stability of the pseudogap phase under uniaxial strain as well as the geometry of its dominant CDW fluctuations. We provide a discussion of the results in Section \ref{sec:Conclusions}.

\section{The Electronic Structure of NbSe$_{2}$\label{sec:NbSe2 band structure}}
The primitive unit cell of NbSe$_{2}$ is shown in Fig.~\ref{fig:NbSe2 structure} to contain two niobium and four selenium atoms. The partially-filled electronic shells include the Se $4p$ orbitals as well as the $4d$ orbitals of the Nb atoms, giving $22$ relevant orbital degrees of freedom per unit cell. A tight binding model for the electronic band structure using these orbitals can be found by solving the Schr\"{o}dinger equation:
\begin{align}
\hat{H} |n\rangle=E_{n} \hat{S} |n\rangle\label{eq:TISE}
\end{align}
where the overlap matrix $S$ is introduced because of the non-orthogonality of orbitals on neighboring atoms. In terms of the local orbital wave functions $|\phi_{n}\rangle$, the operators $\hat{H}$ and $\hat{S}$ are given by the $22 \times 22$ matrices:
\begin{align}
H_{nm} & = \langle\phi_{m}|H|\phi_{n}\rangle \notag \\
S_{nm} & = \langle\phi_{m}|\phi_{n}\rangle
\end{align}
with $H_{nm}$ the hopping amplitudes, and $S_{nm}$ the orbital overlaps.  

\begin{figure}
\begin{centering}
\includegraphics[width=\columnwidth]{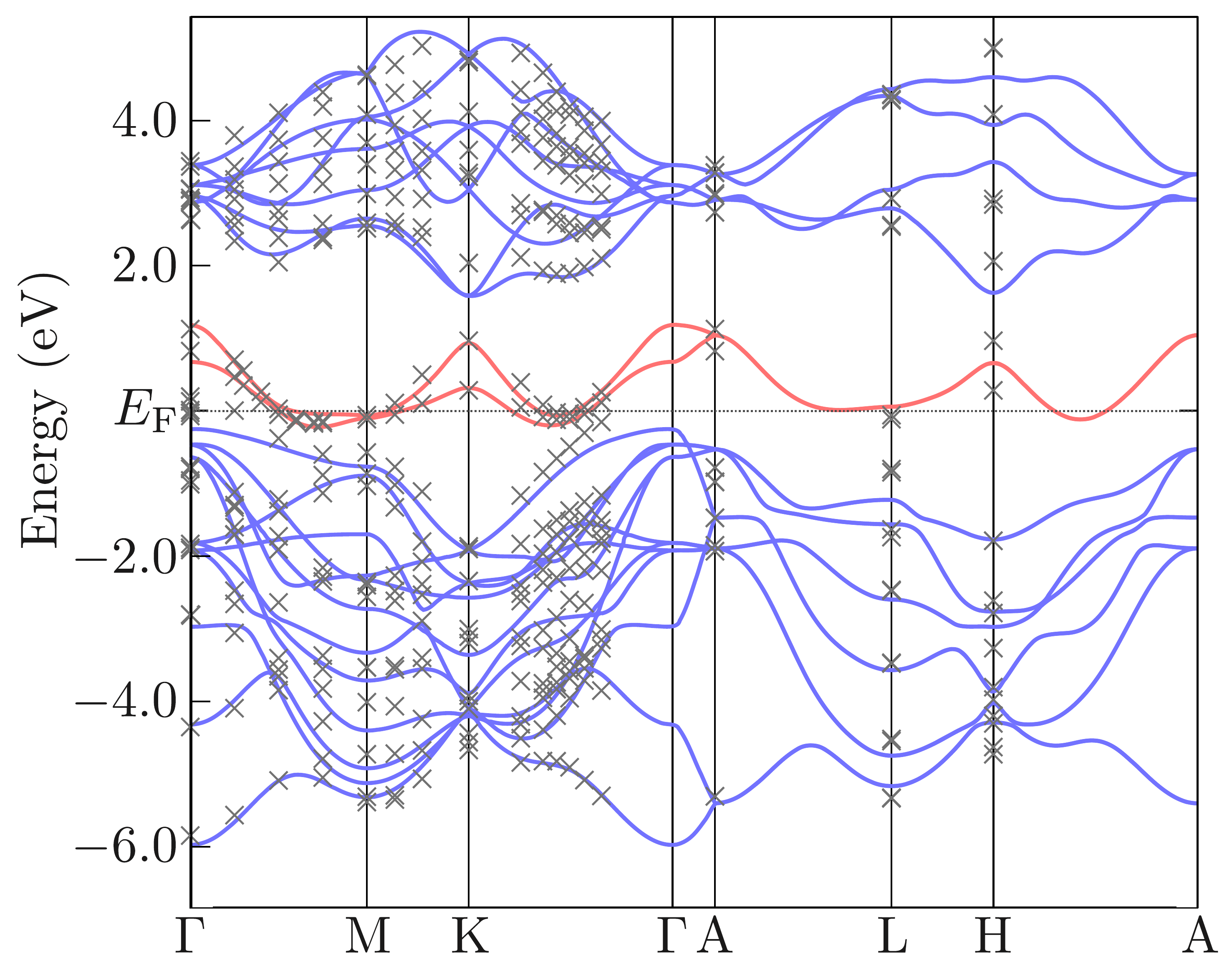}
\end{centering}
\caption{\label{fig:tight binding fit}The result of the $22$-band tight-binding fit to the electronic band structure. The crosses indicate the data being fit to, taken from ARPES for the two red bands crossing $E_{\text{F}}$~\cite{RahnEA12}, and LDA calculations for the remaining $20$ bands in blue~\cite{RossnagelEA01}. The red bands were given larger weight in the fitting procedure.}
\end{figure}

The hopping integrals and overlap integrals are included up to first-nearest neighbors, and related to each other according to the expressions found by Slater and Koster~\cite{KosterSlater54}. Each orbital can additionally be assigned a chemical potential, which is independent from that of orbitals that are different under the imposed lattice symmetries. Altogether, the result is a tight-binding expression with $33$ independent free parameters, which were varied in a Monte Carlo routine to fit two sets of data. First, the ARPES data from Rahn \emph{et al.}~\cite{RahnEA12} provide accurate information about the two bands crossing the Fermi level. For the remaining bands, an earlier Local Density Approximation (LDA) calculation by Rossnagel \emph{et al.}~\cite{RossnagelEA01} was used. The results of the fit are given in Fig.~\ref{fig:tight binding fit}. 

\begin{figure}
\begin{centering}
\includegraphics[width=\columnwidth]{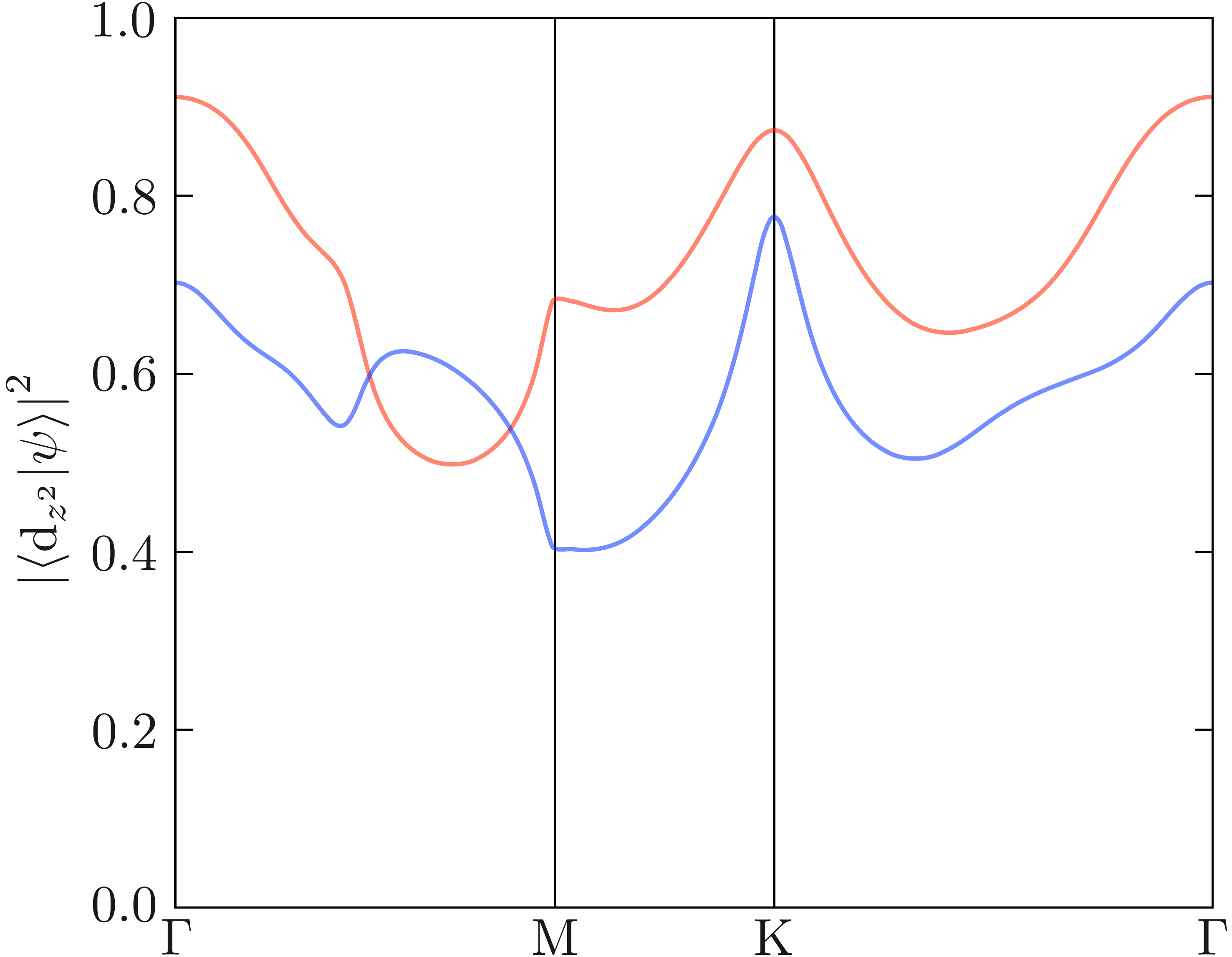}
\end{centering}
\caption{\label{fig:band composition}The contribution of the niobium $d_{3z^{2}-r^{2}}$ orbitals to the two bands crossing $E_{\text{F}}$. The upper, red curve shows the contribution within the lower-energy band making up the inner pocket around the K-point,  while the lower, blue curve shows the same for the higher-energy band or outer pocket. The average contribution of approximately $60\%$ $d_{3z^{2}-r^{2}}$ orbitals across both bands is in agreement with earlier calculations~\cite{Doran78}, and constitutes the dominant orbital character for these bands.}
\end{figure}

The orbital make-up of the bands can be deduced from the eigenvectors found in the tight-binding fitting procedure. The two bands crossing $E_{\text{F}}$ were found to consist primarily of the two niobium $d_{3z^{2}-r^{2}}$ orbitals (one for each layer in the unit cell) throughout the Brillouin zone. As shown in Fig.~\ref{fig:band composition}, these states contribute at least $60\%$ of the orbital character across both bands, in agreement with earlier reports~\cite{Doran78}. 

Projecting the states crossing $E_{\text{F}}$ onto just the two $d_{3z^{2}-r^{2}}$ orbitals, they become exactly equal-weight superpositions, with the lower-energy band being the symmetric combination of the two involved orbitals, and the upper band the anti-symmetric combination. That these superpositions must have precisely equal weights throughout the Brillouin zone is a consequence of the fact that the $d_{3z^{2}-r^{2}}$ orbitals see identical atomic environments up to first-nearest neighbor. This can be made explicit by writing out the tight-binding Schr\"{o}dinger equation~\eqref{eq:TISE} including only the two bands crossing $E_{\text{F}}$:
\begin{align}
\left(\begin{array}{cc} H_{11}-E_{n}S_{11} & \quad H_{12}-E_{n}S_{12}\\ H_{12}-E_{n}S_{12} & \quad H_{11}-E_{n}S_{11} \end{array}\right)\left(\begin{array}{c}\alpha |1\rangle \\ \beta |2\rangle \end{array}\right)=0.
\label{eq:TISE twoband}
\end{align}
Here the states $|1\rangle$ and $|2\rangle$ represent the $d_{3z^{2}-r^{2}}$ orbitals in the upper and lower sandwich layers within a unit cell, so that $H_{11}=H_{22}$ and $S_{11}=S_{22}$. The elements $H_{nm}$ and $S_{nm}$ can be chosen to be real in the absence of time-reversal symmetry breaking. Solving the eigenvalue problem of Eq.~\eqref{eq:TISE twoband} then yields equal-weight superpositions $\alpha=\pm\beta = \sqrt{1/2}$, regardless of any $k$-space structure of either the hopping or overlap integrals.

Including terms up to fifth-nearest neighbor in-plane and second-nearest neighbor out of plane, the two-band tight-binding fit to the bands seen to cross $E_{\text{F}}$ in the ARPES data results in an expression identical to that proposed by Rahn \emph{et al.}~\cite{RahnEA12}.

\section{Electron-Phonon Coupling from the Electronic Bandstructure\label{sec:Varma}}
In order to quantitatively model how the electronic band structure influences the CDW formation and \emph{vice versa}, it is essential to consider both the orbital and momentum dependence of the electron-phonon coupling. This can be done using the expression for the electron-phonon coupling given by Varma \emph{et al.}~\cite{VarmaEA79}, which has been well-tested in the case of transition metal compounds with a predominantly $d$-orbital character at the Fermi level:
\begin{align}
\mathbf{g}_{\mathbf{k},\mathbf{k}'}^{\mu,\nu} \propto \mathbf{v}_{\mathbf{k}}^{\mu}\left[A_{\mathbf{k}}^{\dagger}S_{\mathbf{k}}A_{\mathbf{k}'}\right]^{\mu, \nu} - \left[A_{\mathbf{k}}^{\dagger}S_{\mathbf{k}'}A_{\mathbf{k}'}\right]^{\mu, \nu} \mathbf{v}_{\mathbf{k}'}^{\nu}.\label{eq:g}
\end{align}
Here an overall (purely imaginary) prefactor has been omitted, and $\mathbf{v}^{\mu} = \partial\xi_{\mathbf{k}}^{\mu} / \partial\mathbf{k}$ is the electron velocity in band $\mu$ with dispersion $\xi_{\mathbf{k}}^{\mu}$. The matrix $A_{\mathbf{k}}^{\mu, \nu}$ is the matrix of eigenvectors solving the generalized eigenvalue problem of Eq.~\ref{eq:TISE twoband}, and the vector nature of $\mathbf{g}$ accounts for its coupling to atomic displacements in three real-space directions. The CDW in NbSe$_2$ is known from X-ray diffraction experiments to correspond to the softening of a longitudinal acoustic phonon, with negligible softening of the transverse modes~\cite{WeberEA11,WeberEA13}. Henceforth we consider only the longitudinal part of the electron-phonon coupling.

For the case of NbSe$_{2}$, the two-band tight-binding fit implies the specific forms of the $A$ and $S$ matrices:
\begin{align}
A_{\mathbf{k}}=\frac{1}{\sqrt{2}}\left(\begin{array}{cc}
1 & 1\\
1 & -1
\end{array}\right),
~~~~~~~
S_{\mathbf{k}}=\left(\begin{array}{cc}
\alpha_{\mathbf{k}} & \beta_{\mathbf{k}}\\
\beta_{\mathbf{k}} & \alpha_{\mathbf{k}}
\end{array}\right)
\label{eq:A}
\end{align}
with $\alpha$ and $\beta$ both real. The expression for the electron-phonon coupling therefore simplifies to:
\begin{align}
\mathbf{g}_{\mathbf{k},\mathbf{k}'}^{\pm,\pm} &\propto \left(\alpha_{\mathbf{k}}\pm\beta_{\mathbf{k}}\right)\mathbf{v}_{\mathbf{k}}-\left(\alpha_{\mathbf{k}'}\pm\beta_{\mathbf{k}'}\right)\mathbf{v}_{\mathbf{k}'} \notag \\
\mathbf{g}_{\mathbf{k},\mathbf{k}'}^{\pm,\mp} &=0\label{eq:g simplified}
\end{align}
where the positive sign corresponds to the lower-energy inner pocket and the negative sign to the outer pocket. Across the Brillouin zone, the tight-binding fit indicates that $\beta \approx 2\alpha$, which implies that the size of the electron-phonon coupling is around three times larger in the inner band than the outer. 

To see how this strong orbital dependence of the electron-phonon coupling influences the predicted CDW ordering within the tight-binding model of the electronic structure, consider the general action describing coupled electrons and phonons:
\begin{align}
S  =& \sum_{k\nu}\psi_{k,\nu}^{\dagger}G_{k,\nu}^{-1}\psi_{k,\nu}^{\phantom{\dagger}}+\sum_{q}\varphi_{q}^{\dagger}D_{q}^{-1}\varphi_{q}^{\phantom{\dagger}}\notag\\
&+\sum_{kq\mu\nu}g_{\mathbf{k},\mathbf{k}+\mathbf{q}}^{\mu, \nu}\varphi_{q}^{\phantom{\dagger}}\psi_{k+q,\nu}^{\dagger}\psi_{k,\mu}^{\phantom{\dagger}}.
\label{eq:Sfull}
\end{align}
Here $\mu$ and $\nu$ are electronic band indices, while $G_{k,\nu}$ and $D_{q}$ are propagators for the electron field $\psi_{k,\nu}$ and phonon field $\varphi_{q}$ respectively. They can be written in the standard forms:
\begin{align}
G_{k,\nu} &= \left(i\omega_{n}-\xi_{\mathbf{k}}^{\nu}+\mu\right)^{-1}\notag \\
D_{q} &=\frac{-2\Omega_{\mathbf{q}}}{\left(i\Omega_{n}\right)^{2}-\Omega_{\mathbf{q}}^{2}}
\end{align}
with fermionic Matsubara frequencies $i \omega_{n}$, and bosonic frequencies $i \Omega_{n}$. The bare phonon dispersion is $\Omega_{\mathbf{q}}$. 

Integrating out the electrons from Eq.~\eqref{eq:Sfull} yields an effective action, which, to quadratic order in the phonon fields, is given by:
\begin{align}
S_{\text{eff}}\left[\varphi\right] =&\sum_{q}\varphi_{q}^{\dagger}\left(D_{q}^{-1}+\frac{1}{2}D_{2}\left(\mathbf{q}\right) \right)\varphi_{q}^{\phantom{\dagger}} \label{eq:Seff}
\end{align}
where $D_2\left(\mathbf{q}\right)$ is the experimentally-accessible generalized static electronic susceptibility, defined as:
\begin{align}
D_{2}\left(\mathbf{q}\right) = -\sum_{\mathbf{k},\mu,\nu}g_{\mathbf{k},\mathbf{k}+\mathbf{q}}^{\mu, \nu}g_{\mathbf{k}+\mathbf{q},\mathbf{k}}^{\nu, \mu}\frac{f(\xi_{\mathbf{k}}^{\mu})-f(\xi_{\mathbf{k}+\mathbf{q}}^{\nu})}{\xi_{\mathbf{k}}^{\mu}-\xi_{\mathbf{k}+\mathbf{q}}^{\nu}}\label{eq:D2}
\end{align}
where $f(\xi_{\mathbf{k}}^{\mu})$ is the Fermi-Dirac distribution function. Notice that this expression reduces to the more-commonly encountered bare electronic susceptibility $\chi\left(\mathbf{q}\right)$ if $g_{\mathbf{k},\mathbf{k}+\mathbf{q}}$ is approximated to be independent of momentum. 
\begin{figure}
\begin{centering}
\includegraphics[width=\columnwidth]{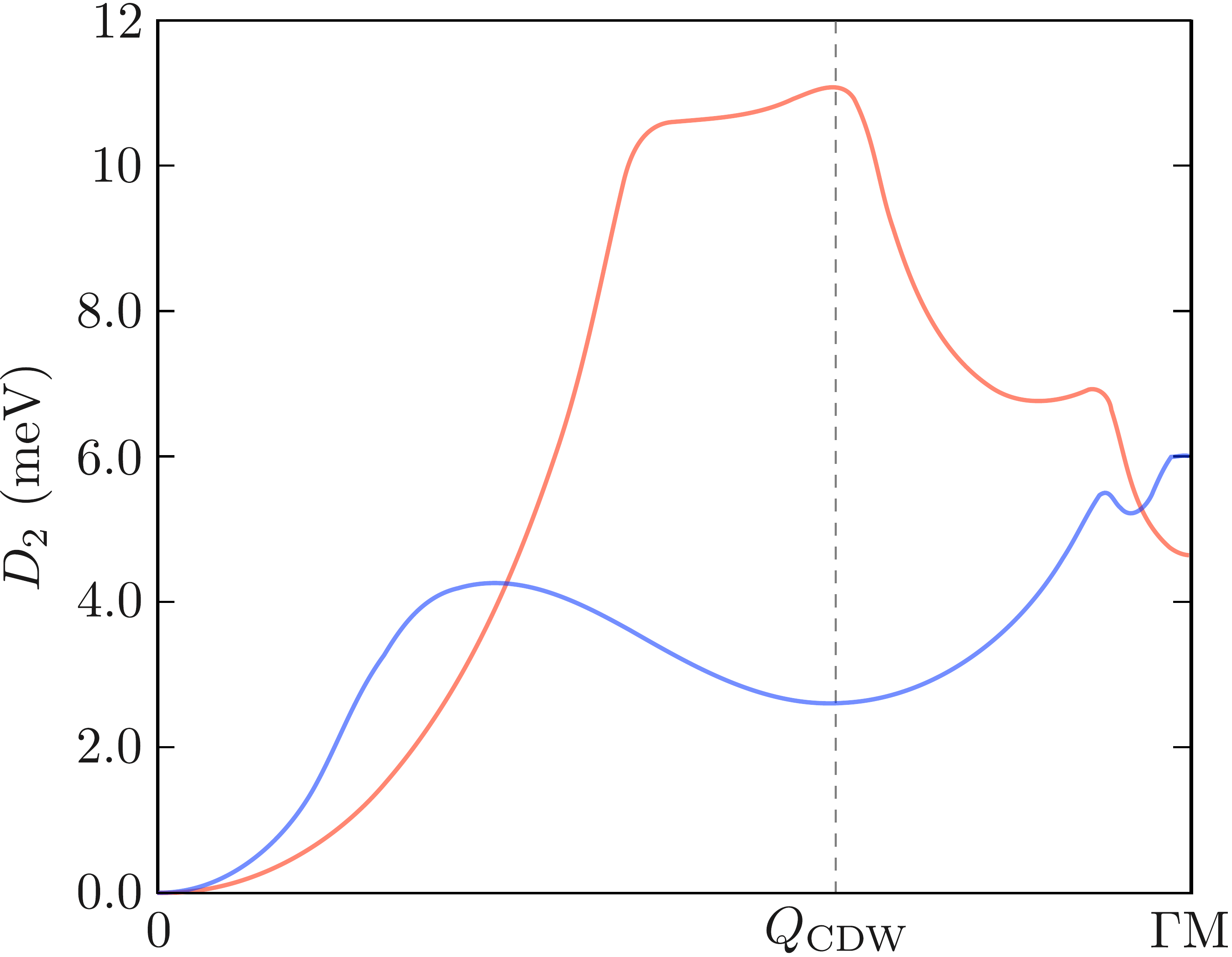}
\end{centering}
\caption{\label{fig:D2 (alpha, g)}The intra-band contributions to the generalized static electronic susceptibility $D_{2}$ originating in the two bands crossing $E_{\text{F}}$. The upper, red, curve corresponds to the lower-energy band making up the inner pocket around the K-point, in which the CDW gap has been found experimentally. The magnitude of the electron-phonon coupling is set to give the mean-field (RPA) phase transition at $33.5\,$K, and the Fermi-Dirac distributions are evaluated at the same temperature.}
\end{figure}

Figure~\ref{fig:D2 (alpha, g)} compares the intra-band contributions to the generalized susceptibility from the two bands crossing $E_{\text{F}}$ in NbSe$_2$. Since the inter-band electron-phonon coupling was found in Eq.~\eqref{eq:g simplified} to be zero, the two intra-band contributions can be thought of as two independent electronic susceptibilities, each operative in its own electronic band. Figure~\ref{fig:D2 (alpha, g)} then shows that the orbital and momentum dependence of the electron-phonon coupling results in a susceptibility that is at least three times larger in the lower-energy band (the inner pocket) than in the higher-energy band (the outer pocket) at values of the momentum transfer corresponding to the CDW wave vector. Since the CDW gap size is proportional to the susceptibility in each band, the orbital and momentum dependence of the electron-phonon coupling leads to large differences in gap magnitudes for the two bands crossing $E_{\text{F}}$, and provides an explanation for the experimental observation that the CDW gap is confined primarily to a single band~\cite{BorisenkoEA09,RahnEA12}.
\begin{figure}
\begin{centering}
\includegraphics[width=\columnwidth]{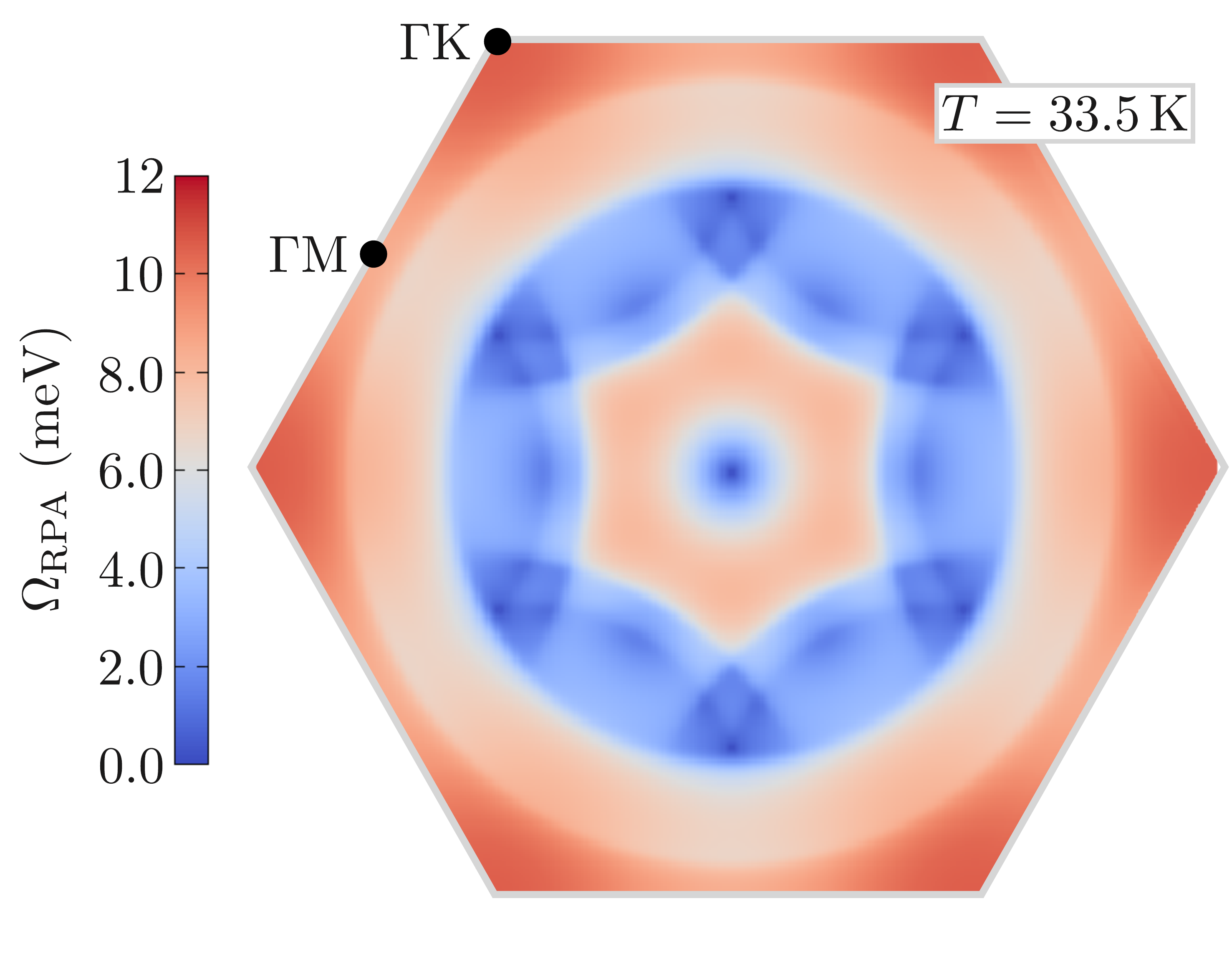}
\end{centering}
\caption{\label{fig:Omega_RPA}The RPA-renormalized phonon frequency as a function of momentum transfer $\mathbf{q}$. The magnitude of the electron-phonon coupling has been set to give the RPA phase transition at $33.5\,$K, and the phonon frequencies shown are evaluated at that critical temperature.}
\end{figure}

Returning to the expression for the effective action in Eq.~\eqref{eq:Seff}, the term in parentheses can be interpreted to represent the inverse propagator for phonons renormalized by the presence of electron-phonon interactions. The renormalized  phonon frequency is then written as:
\begin{align}
\Omega_{\text{RPA}}^{2}\left(\mathbf{q}\right)=\Omega_{\mathbf{q}}^{2}-\Omega_{\mathbf{q}}D_{2}\left(\mathbf{q},\Omega\right).
\label{eq:omega_D2}
\end{align}
The label `RPA' indicates that this constitutes the Random Phase Approximation or mean-field form of the renormalized phonon energy. As temperature is lowered, the generalized susceptibility becomes more-sharply peaked, giving a more-pronounced dip in the renormalized phonon dispersion. Once $\Omega_{\text{RPA}}$ touches zero energy at a given momentum, the atomic structure becomes unstable and a CDW develops at the wave vector corresponding to the momentum of the soft mode. 

\begin{figure}
\begin{centering}
\includegraphics[width=\columnwidth]{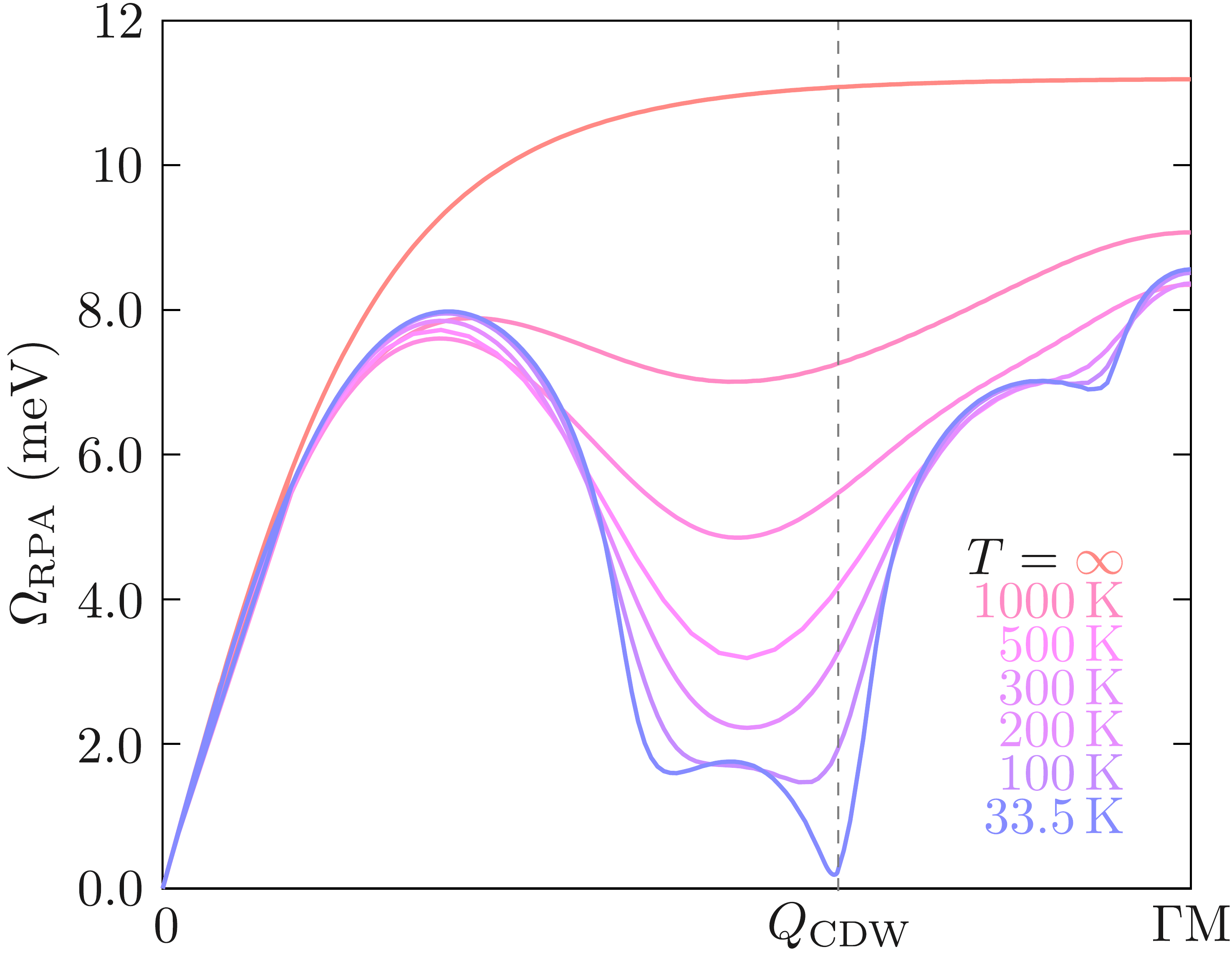}
\end{centering}
\caption{\label{fig:Omega_RPAb}The phonon dispersion along the high-symmetry direction $\Gamma$M at various temperatures. The magnitude of the electron-phonon coupling has been set to give the RPA phase transition at $33.5\,$K. The $T=\infty$ curve shows the bare phonon dispersion. At $T_{\text{RPA}}=T_{\text{CDW}}=33.5\,$K the soft phonon mode can be seen to select out the experimentally-observed wavector $\mathbf{Q}_{\text{CDW}}$.}
\end{figure}

To see what wave vector is predicted for the CDW order in NbSe$_2$ by the present tight-binding model, we employ a phenomenological fit to X-ray diffraction experiments for the bare phonon frequency $\Omega_{\mathbf{q}}$~\cite{WeberEA13}. Figure~\ref{fig:Omega_RPA} shows the resulting renormalized phonon frequency, when setting the overall magnitude $g$ of the electron-phonon coupling such that a CDW instability develops at the experimentally-observed transition temperature of  $33.5\,\mbox{K}$. The plot across the Brillouin zone shows that the phonon mode first softens to zero along $\Gamma$M. The momentum-space cut in this direction, whose evolution with temperature is displayed in figure~\ref{fig:Omega_RPAb}, reveals that the instability in fact occurs precisely at the known ordering vector of the 3Q CDW state, as seen for example in neutron scattering~\cite{MonctonEA75,MonctonEA77}, or X-ray diffraction experiments~\cite{WeberEA11, WeberEA13, FengEA15}. The presence of a broad plateau of partially-softened phonon frequencies surrounding the CDW wave vector, similar to that seen experimentally~\cite{WeberEA11, WeberEA13}, is a direct result of the strong electron-phonon coupling in NbSe$_2$, or, equivalently, of the absence of a truly nested electronic structure.

\section{The Extent of Nesting in NbSe$_{2}$ \label{sec:nesting}}
To quantify the extent to which electron-phonon coupling, as opposed to nesting of the electronic structure, drives the CDW transition in NbSe$_2$, it is convenient to consider temporarily a simplified form of the electron-phonon coupling, which is a function only of the magnitude of the momentum transfer: $g_{\mathbf{k},\mathbf{k}+\mathbf{q}}\rightarrow g_{\left|\mathbf{q}\right|}$. In this case, the generalized susceptibility is simply the product of the square of the electron-phonon coupling with the bare electronic susceptibility: 
\begin{align}
D_{2}\left(\mathbf{q}\right) &= \sum_{\mu,\nu} \left( g^{\mu, \nu}_{\left|\mathbf{q}\right|} \right)^2 \chi^{\mu, \nu}\left(\mathbf{q}\right) \notag \\
\chi^{\mu, \nu}\left(\mathbf{q}\right) &= -\sum_{\mathbf{k}} \frac{f(\xi_{\mathbf{k}}^{\mu})-f(\xi_{\mathbf{k}+\mathbf{q}}^{\nu})}{\xi_{\mathbf{k}}^{\mu}-\xi_{\mathbf{k}+\mathbf{q}}^{\nu}}.\label{eq:chi}
\end{align}
In this expression it is clear that the denominator in the susceptibility $\chi$ causes a divergence whenever the Fermi surface is nested.
\begin{figure}
\begin{centering}
\includegraphics[width=\columnwidth]{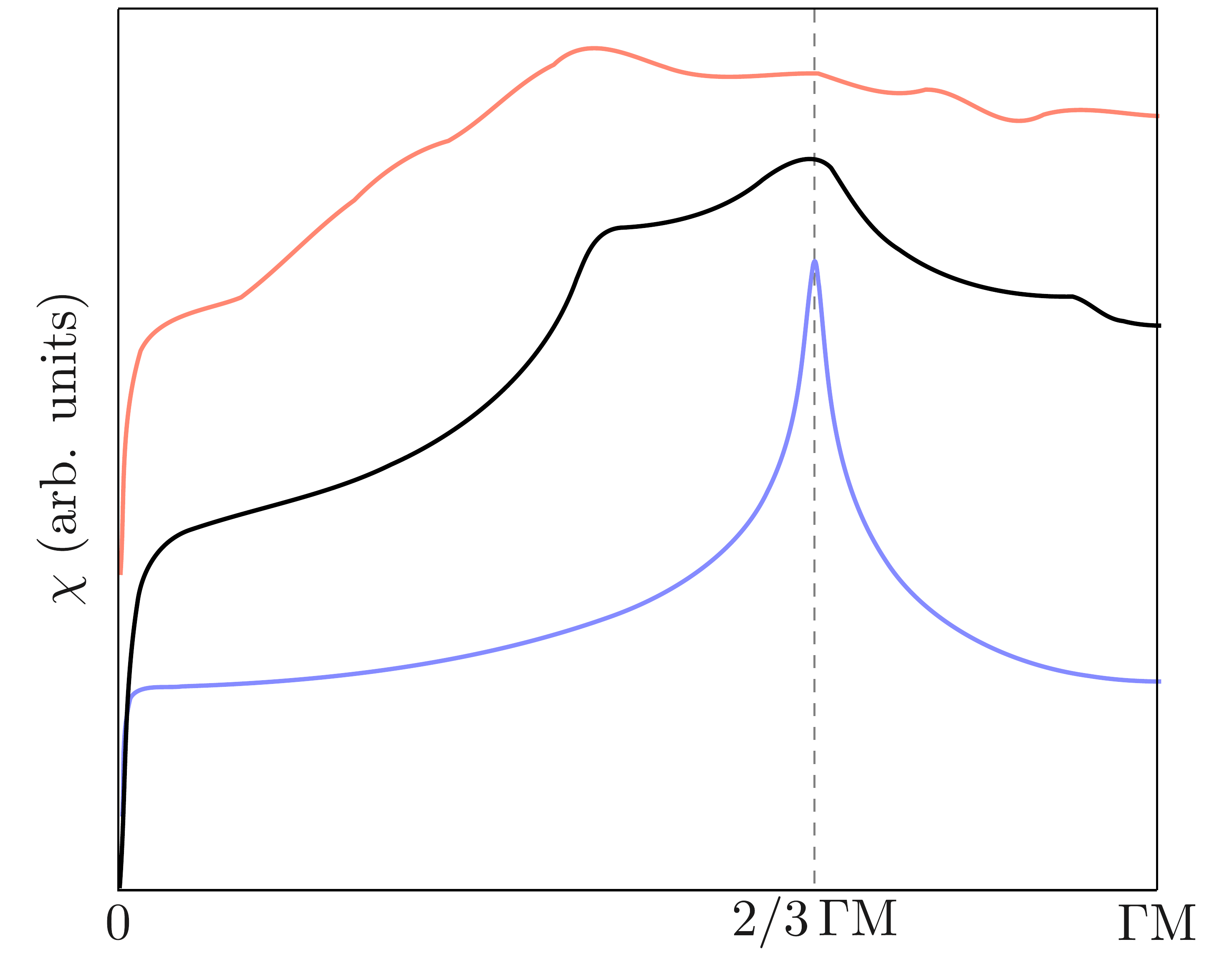}
\end{centering}
\caption{\label{fig:susceptibility}The bare electronic susceptibility, $\chi(\mathbf{q})$, for three special cases. Blue: the susceptibility resulting from a single 1D band with cosine dispersion and nesting vector $\frac{2}{3}{\Gamma}$M. Red: the sum of susceptibilities in the $2$-band fit to the NbSe$_{2}$ band structure with constant and equally-weighted interband and intraband scattering. This corresponds to the generalized susceptibility $D_2(\mathbf{q})$ if both the momentum and orbital dependence of the electron-phonon coupling are ignored. Black: the bare susceptibility of only the single NbSe$_{2}$ band developing CDW gaps. In all cases a temperature of $T=33.5\,\mbox{K}$ was used along with a $5\,\mbox{meV}$ numerical regularization in the expression for the bare susceptibility. The curves have been offset vertically relative to each other for clarity.}
\end{figure}

The bare susceptibility $\chi$ defined by Eq.~\eqref{eq:chi} is shown in Fig.~\ref{fig:susceptibility} for three special cases. The bottom (blue) curve is based on a prototypical well-nested one-dimensional dispersion with a single band. The upper (red) curve employs the tight-binding fit to the two bands crossing $E_{\text{F}}$ in NbSe$_2$, and shows the sum of both intraband and interband susceptibilities. It corresponds to a generalized susceptibility for NbSe$_2$ if equally-weighted interband and intraband couplings are artificially imposed ($g^{\mu, \nu}_{\left|\mathbf{q}\right|} \equiv 1$). The central (black) curve is the bare susceptibility for the lower-energy NbSe$_{2}$ band only. The final case is expected to most-closely approximate the actual electronic susceptibility of NbSe$_{2}$, as the orbital dependence of the electron-phonon coupling was shown in the previous section to be dominated by the intraband contribution from the inner electron pocket. The case of equally-weighted interband and intraband susceptibilities has been extensively used in previous studies of NbSe$_2$~\cite{JohannesEA06,RahnEA12,SoumyanarayananEA13}.

The nested 1D band structure can be seen in Fig.~\ref{fig:susceptibility} to yield a true divergence of the susceptibility at the nesting vector, which in this case was chosen to be $\frac{2}{3}\Gamma$M. Both band structures based on the NbSe$_2$ tight-binding model, on the other hand, give rise to rather flat susceptibilities. Ignoring the orbital dependence of the electron-phonon coupling, and taking an equal-weight sum of interband and intraband susceptibilities, yields a total susceptibility with a maximum at a momentum value which clearly differs from the observed CDW ordering vector. The susceptibility arising from just the inner band peaks at the correct momentum, but compared to the true divergence of the 1D case, NbSe$_{2}$ can qualitatively be said to lack nesting in its electronic structure. 

This statement can be made quantitative by considering how the predicted CDW ordering vector varies with changes in the electron-phonon coupling. Within the approximation that the electron-phonon coupling depends only on the transferred momentum, the renormalized phonon frequency of Eq.~\eqref{eq:omega_D2} becomes:
\begin{align}
\Omega_{\text{RPA}}^{2} &=\Omega_{\mathbf{q}}^{2}-\Omega_{\mathbf{q}}\sum_{\mu, \nu}\left(g^{\mu, \nu}_{\left|\mathbf{q}\right|}\right)^{2}\chi^{\mu, \nu}\left(\mathbf{q},\Omega\right).
\label{eq:omega_RPA}
\end{align}

Given the flat electronic susceptibility $\chi$ of Fig.~\ref{fig:susceptibility}, the experimentally-observed softening of the phonon in NbSe$_2$~\cite{WeberEA11,WeberEA13} can only be reproduced by $\Omega_{\text{RPA}}$ if the electron-phonon coupling $g_{\mathbf{\left|q\right|}}$ is itself peaked at the observed CDW wave vector. In fact, Eq.~\eqref{eq:omega_RPA} can be used to deduce the dependence of the electron-phonon coupling on momentum transfer, given the dispersions of both the softened and bare phonons.

\begin{figure}
\begin{centering}
\includegraphics[width=\columnwidth]{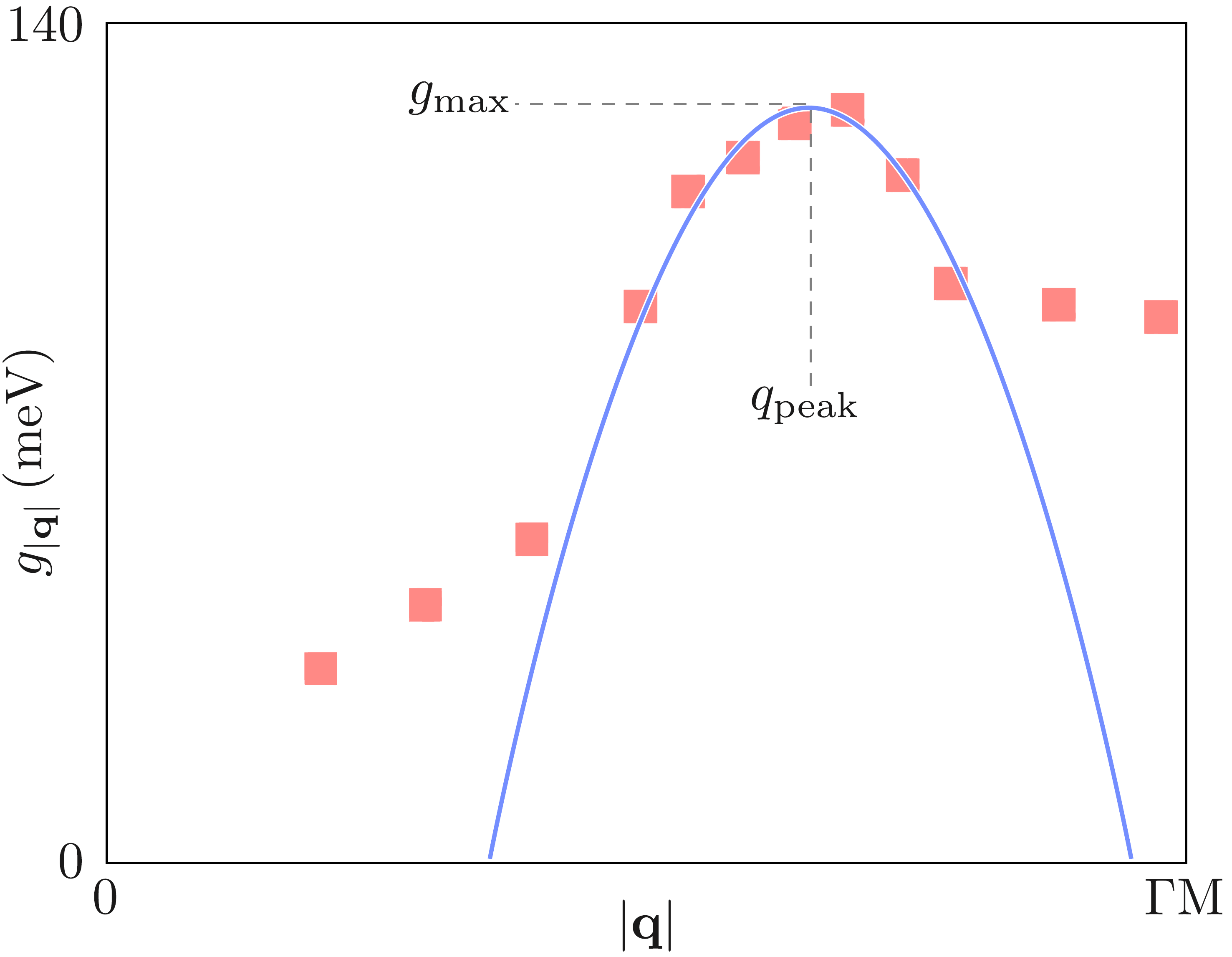}
\end{centering}
\caption{\label{fig:g estimate}Least-squares parabolic fit to the simplified electron-phonon coupling $g_{|\mathbf{q}|}$ around $\mathbf{Q}_{\text{CDW}}$. The red data points are based on the bare and renormalized phonon frequencies measured by inelastic x-ray scattering~\cite{WeberEA13}, combined with the bare electronic susceptibility, as described in the text.}
\end{figure}

The bare phonon dispersion can be approximated by that measured experimentally at high temperatures, while a maximally-renormalized phonon dispersion is realized at the CDW transition temperature~\cite{WeberEA11,WeberEA13}. Together with the temperature-dependent susceptibility $\chi$ calculated from the electronic band structure, this results in the form of the electron-phonon coupling $g_{\left|\mathbf{q}\right|}$ shown in Fig.~\ref{fig:g estimate}. Around its maximum, the obtained momentum dependence is well-described by a parabolic fit of the form:
\begin{align}
g_{\left|\mathbf{q}\right|}=-a\left(1-\frac{\left|\mathbf{q}\right|}{q_{\text{peak}}}\right)^{2}+g_{\text{max}}.\label{eq:g fit}
\end{align}
The best-fit values $g_{\text{max}}=132\,\mbox{meV}$, and $a=628\,\mbox{meV}$ were used, while $q_{\text{peak}}$ was restricted to the experimentally-observed value of $q_{\text{peak}}=0.657\left|{\Gamma\text{M}}\right|$~\cite{FengEA15}. The electron-phonon coupling obtained in this way is approximately independent of temperature.

Given a specific functional form for the electron-phonon coupling $g_{\left|\mathbf{q}\right|}$, the CDW ordering vector, as predicted by Eq.~\eqref{eq:omega_RPA}, is the momentum transfer at which the renormalized phonon dispersion $\Omega_{\text{RPA}}$ first touches zero energy. This fact can now be used to quantify the amount of nesting present in the band structure of NbSe$_2$ by considering the change of the predicted CDW wave vector $\mathbf{Q}_{\text{CDW}}$ as the electron-phonon coupling's peak momentum $q_{\text{peak}}$ is varied.

In a truly nested system, the bare electronic susceptibility has a divergence, which dominates the convolution of $g_{\left|\mathbf{q}\right|}$ and $\chi$ in Eq.~\eqref{eq:omega_RPA} and which determines the value of $\mathbf{Q}_{\text{CDW}}$ independent of the momentum dependence of the electron-phonon coupling. Conversely, in the presence of a band structure without any nesting, the bare electronic susceptibility will be an approximately-flat function of momentum, and the predicted CDW wave vector $\mathbf{Q}_{\text{CDW}}$ will be determined entirely by the shape of the electron-phonon coupling.
\begin{figure}
\begin{centering}
\includegraphics[width=\columnwidth]{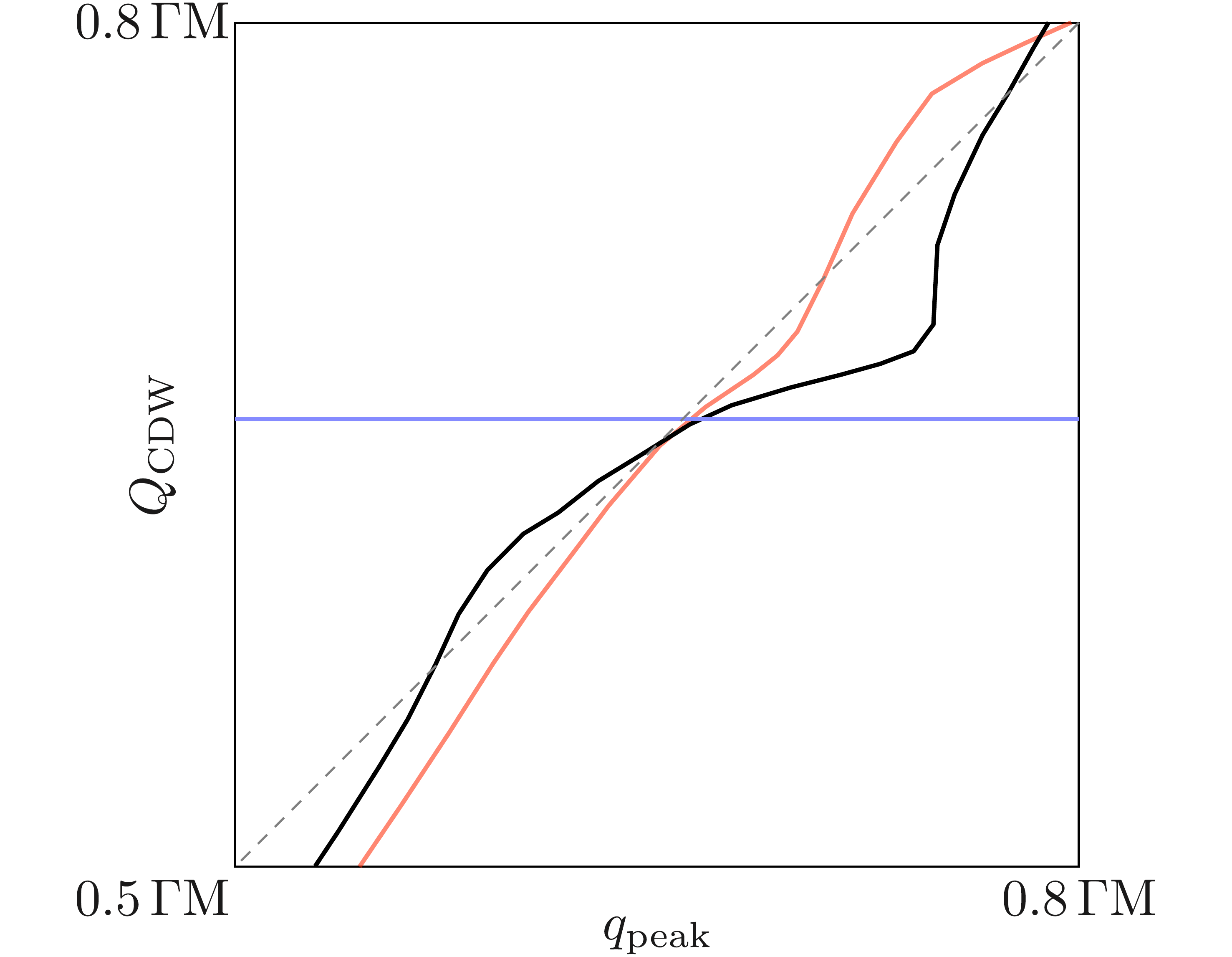}
\end{centering}
\caption{\label{fig:no nesting}The CDW ordering vector $Q_{\text{CDW}}$ obtained from the mean-field renormalized phonon dispersion $\Omega_{\text{RPA}}$, assuming that the electron-phonon coupling can be described by the simplified form $g_{|\mathbf{q}|}$, with a variable peak position. Blue: the ordering vector for a 1D band structure with nesting vector $\frac{2}{3}\Gamma$M. Red: the ordering vector for the two-band tight-binding fit to the NbSe$_{2}$ electronic structure, imposing that the bare susceptibility is an equal-weight sum of interband and intraband contributions. Black: the ordering vector based on the bare susceptibility of only the lower-energy band crossing $E_{\text{F}}$ in the NbSe$_{2}$ band structure. The susceptibilities corresponding to all three cases are shown in Fig.~\ref{fig:susceptibility}. The grey dots depict the line $Q_{\text{CDW}}=q_{\text{peak}}$. }
\end{figure}

Taking the parabolic fit to the electron-phonon coupling found in Fig.~\ref{fig:g estimate} and varying the value of the peak position, the dependence of $Q_{\text{CDW}}$ on $q_{\text{peak}}$, as predicted by the softening of $\Omega_{\text{RPA}}$, can be mapped out. The electronic susceptibility can again be taken from either a well-nested band structure, from an equally-weighted sum of interband and intraband contributions in the two-band tight-binding model for NbSe$_2$, or from the single band in the NbSe$_2$ tight-binding structure which dominates its electron-phonon coupling. In the case of a one-dimensional band structure, the resulting prediction for $\mathbf{Q}_{\text{CDW}}$ is the flat blue line in Fig.~\ref{fig:no nesting}, reflecting the fact that one-dimensional systems are always well-nested. If instead the total susceptibility based on equally-weighted interband and intraband contributions from the tight-binding band structure of NbSe$_2$ is employed, the result is the red curve in Fig.~\ref{fig:no nesting}. Around $\mathbf{Q}_{\text{CDW}}=\frac{2}{3}\Gamma$M it closely follows the diagonal line $Q_{\text{CDW}}=q_{\text{peak}}$ (where $Q_{\text{CDW}}=|\mathbf{Q}_{\text{CDW}}|$) indicating that the susceptibility has little momentum dependence, and the electron-phonon coupling dominates in selecting the CDW propagation vector. Finally, using the tight-binding band structure of NbSe$_2$, but including only the lowest-energy band crossing $E_{\text{F}}$, the ordering vector depends on the electron-phonon coupling as shown by the black line in Fig.~\ref{fig:no nesting}, and lies in between the previous two extreme cases. 

The quantity $\alpha = 1 - \partial Q_{\text{CDW}} / \partial q_{\text{peak}}$, evaluated at the point where $q_{\text{peak}}$ equals the experimentally-observed CDW wave vector, can be used to quantify the influence of nesting as compared to that of electron-phonon coupling. It equals one for the perfectly-nested one-dimensional system, and zero in the case of a perfectly momentum-independent electronic susceptibility. The orbital dependence of the electron-phonon coupling was shown in the previous section to lead to a dominant size of the coupling within the lowest-energy band. It may therefore be expected that, within the present approximation, the black curve in Fig.~\ref{fig:no nesting}, taking into account only the lowest-energy band, is the most relevant representation of the bare electronic susceptibility in NbSe$_2$. It is characterized by the nesting parameter $\alpha = 0.55$, indicating that neither the electronic structure nor the momentum dependence of the electron-phonon coupling can be neglected.

Having used the reduced form $g_{\left|\mathbf{q}\right|}$ to investigate the extent of nesting, we now return to the full expression $g_{\mathbf{k},\mathbf{k}+\mathbf{q}}^{\mu, \nu}$ of Eq.~\eqref{eq:g} for the remainder of this article.

\section{The CDW Gap\label{sec:The-CDW-Gap}}
To obtain a prediction for the momentum and energy dependence of the CDW gap in NbSe$_2$ based on the electronic tight-binding model and the momentum- and orbital-resolved electron-phonon coupling, we now establish a gap equation in line with the Bogoliubov-de Gennes philosophy. In contrast to its more familiar use within the theory of superconductivity, the gap matrix for CDW order is written in a basis of operators at different values of the momentum rather than of particle and hole operators. The reason is that a charge ordered pattern with propagation vector $\mathbf{Q}_{\text{CDW}}$ enlarges the real space unit cell of its host material, and correspondingly reduces the size of its first Brillouin zone, making states with momenta $\mathbf{k}$ and $\mathbf{k}+m \mathbf{Q}_{\text{CDW}}$ equivalent ($m$ integer). In the case of a multi-$Q$ CDW with multiple ordering vectors $\mathbf{Q}^j_{\text{CDW}}$, the equivalence even extends to states at $\mathbf{k}+\sum_j m_j \mathbf{Q}^j_{\text{CDW}}$, with $m_j$ a set of independent integers. This results in a basis for the Bogoliubov-de Gennes gap equation containing different entries for all distinct momentum values in the original Brillouin zone which become equivalent in the reduced Brillouin zone.
\begin{figure}
\begin{centering}
\includegraphics[width=\columnwidth]{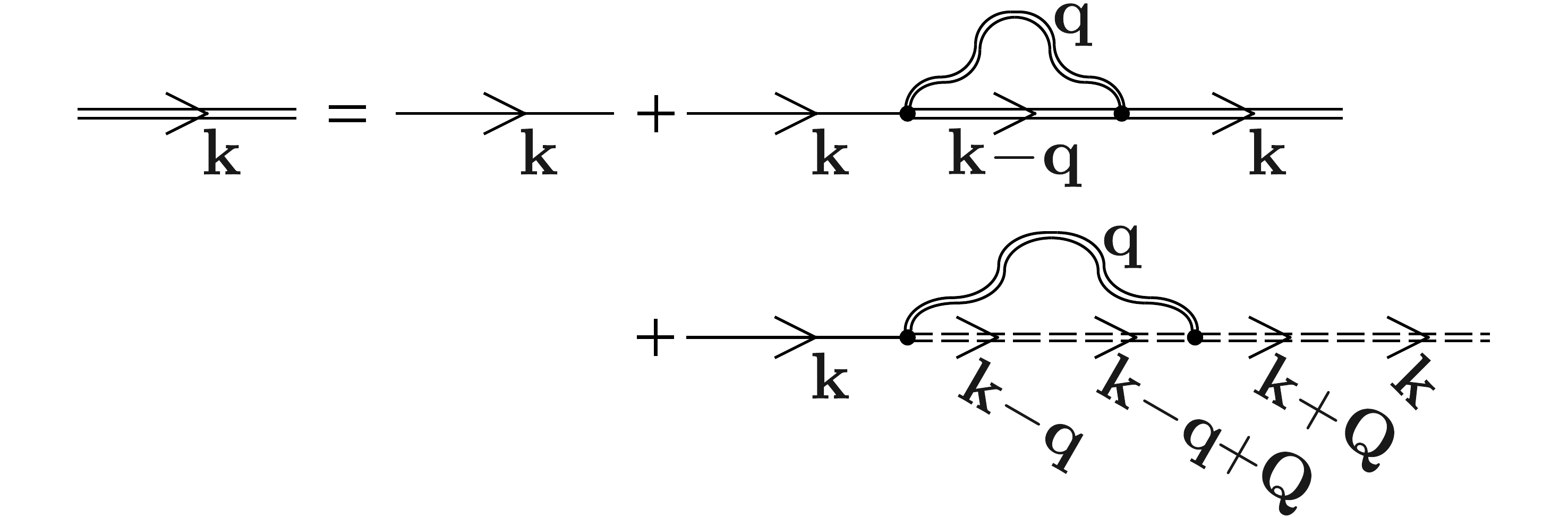}
\end{centering}
\caption{\label{fig:Feynman}Diagrammatic depiction of the Bogoliubov-de Gennes gap equation. Double straight lines indicate self-consistently renormalized electronic propagators, while double wavy lines represent phonon propagators within the mean-field (RPA) approximation. Double dashed lines depict the anomalous Green's functions, which change the crystal momentum by a CDW wave vector. The self energy contribution in the central part of the top right diagram corresponds to a diagonal entry in the matrix of Eq.~\eqref{eq:Sigma_matrix}, while the gap functions for the off-diagonal entries correspond to the central part of the bottom diagram.}
\end{figure}

In the case of NbSe$_2$ the experimentally-observed CDW wave vector is incommensurate, and close to $\mathbf{Q}_{\text{CDW}} = 0.657 \,\Gamma$M. Strictly speaking, this implies that there are infinitely many momentum values in the original reduced Brillouin zone which become equivalent under addition of the CDW ordering vector. As can be seen in STM experiments on NbSe$_2$~\cite{SoumyanarayananEA13}, however, the CDW order is actually organized in patches which are locally commensurate with the underlying lattice, and have local ordering vectors $\mathbf{Q}_{\text{CDW}} = 2/3 \, \Gamma$M. We therefore consider a Bogoliubov-de Gennes gap matrix in a basis of momentum values $\mathbf{k}$, $\mathbf{k}\pm\mathbf{Q}_1$, $\mathbf{k}\pm\mathbf{Q}_2$, $\mathbf{k}\pm\mathbf{Q}_3$, and $\mathbf{k} \pm\left(\mathbf{Q}_{1}-\mathbf{Q}_{2}\right)$. This set of nine entries encapsulates all distinct momentum values in the original Brillouin zone which are equivalent to the point $\mathbf{k}$ in the reduced Brillouin zone. The Bogoliubov-de Gennes equation is given diagramatically in Fig.~\ref{fig:Feynman}, which corresponds to the gap matrix:
\begin{align}
\mathbf{\Sigma} = 
\left(\begin{array}{ccccccccc}
\Sigma_{k} & \Delta & \Delta & \Delta & \Delta & \Delta & \Delta & 0 & 0\\
\Delta & \Sigma_{1} & \Delta & 0 & \Delta & 0 & \Delta & \Delta & \Delta\\
\Delta & \Delta & \Sigma_{\bar{1}} & \Delta & 0 & \Delta & 0 & \Delta & \Delta\\
\Delta & 0 & \Delta & \Sigma_{2} & \Delta & 0 & \Delta & \Delta & \Delta\\
\Delta & \Delta & 0 & \Delta & \Sigma_{\bar{2}} & \Delta & 0 & \Delta & \Delta\\
\Delta & 0 & \Delta & 0 & \Delta & \Sigma_{3} & \Delta & \Delta & \Delta\\
\Delta & \Delta & 0 & \Delta & 0 & \Delta & \Sigma_{\bar{3}} & \Delta & \Delta\\
0 & \Delta & \Delta & \Delta & \Delta & \Delta & \Delta & \Sigma_{1\bar{2}} & 0\\
0 & \Delta & \Delta & \Delta & \Delta & \Delta & \Delta & 0 & \Sigma_{\bar{1}2}
\end{array}\right).\label{eq:Sigma_matrix}
\end{align}
The indices on the diagonal elements indicate the momentum value at which the entry is to be evaluated. The element $\Sigma_{1\bar{2}}$, for example, represents the self energy function $\Sigma(\mathbf{k}+\mathbf{Q}_1-\mathbf{Q}_2)$. Each $\Delta$ in the expression should be considered to have two indices, corresponding to the row and column labels, which indicate the momentum scattered from and to. For example, entry $(2,3)$ represents the gap function $\Delta^{\mathbf{k}+\mathbf{Q}_1}_{\mathbf{k}-\mathbf{Q}_1}$.

To express the gap and self energy functions in terms of electron and phonon propagators, it is convenient to employ the so-called Lehmann representation. In this formalism, element $\left(n,m\right)$ of the Bogoliubov-de Gennes gap matrix is written as:
\begin{widetext}
\begin{align}
\mathbf{\Sigma}_{m}^{n}\left(\mathbf{k},\epsilon+i\delta\right) = & -\frac{1}{\pi}\sum_{\mathbf{q}}\left(g_{\mathbf{k}+\mathbf{Q}_{n},\mathbf{k}-\mathbf{q}+\mathbf{Q}_{n}}g_{\mathbf{k}-\mathbf{q}+\mathbf{Q}_{m},\mathbf{k}+\mathbf{Q}_{m}}\left(\frac{\Omega_{\mathbf{q}}}{\Omega_{\text{RPA}}}\right)\int\mbox{d}\epsilon'\mathfrak{Im}\left[G_{m}^{n}\left(\mathbf{k}-\mathbf{q},\epsilon'+i\delta\right)\right] \right. \notag \\
&\left. \cdot\left\{\frac{n_{B}\left(\Omega_{\text{RPA}}\right)+1-f\left(\epsilon'\right)}{\epsilon-\epsilon'-\Omega_{\text{RPA}}+i\delta}+\frac{n_{B}\left(\Omega_{\text{RPA}}\right)+f\left(\epsilon'\right)}{\epsilon-\epsilon'+\Omega_{\text{RPA}}+i\delta}\right\} \right)\label{eq:gap_matrix}
\end{align}
with the corresponding Feynman diagrams contained in Fig.~\ref{fig:Feynman}. The generalized susceptibility in this case reads:
\begin{align}
D_{2}=-\frac{1}{\pi}\sum_{\mathbf{k}}\left|g_{\mathbf{k},\mathbf{k}+\mathbf{q}}\right|^{2}\int\mbox{d}\epsilon'\mathfrak{Im}\left[G^{1}_{1}\left(\mathbf{k},\epsilon'+i\delta\right)\right]\mathfrak{Re}\left[\frac{f(\epsilon')-f(\xi_{\mathbf{k}-\mathbf{q}})}{\epsilon'-\xi_{\mathbf{k}-\mathbf{q}}-i\Omega}\right].
\end{align}
\end{widetext}

For the remainder of this section, we neglect the diagonal (self energy) terms in the matrix of Eq.~\ref{eq:Sigma_matrix}, so that the calculation effectively takes place in the random phase approximation, and the gap equation depicted in Fig.~\ref{fig:Feynman} can in principle be solved self-consistently. To simplify the calculation, we assume the CDW gap to be static, and only consider its value at $\epsilon=0$. We furthermore assume $\Delta_{k+Q_{m}}^{k}=\Delta_{k-Q_{m}}^{k}=\Delta_{k+Q_{n}}^{k}$, for all $m,n \in \{1,2,3\}$, which we numerically confirmed to be a valid approximation within the first iteration. Describing all these elements with the single gap function $\Delta(\mathbf{k})$ allows the remaining elements to be expressed by different instances of the same function, so that for example $\Delta_{k-Q_{1}}^{k+Q_{1}} = \Delta_{k+2Q_{1}}^{k+Q_{1}} = \Delta(\mathbf{k}+\mathbf{Q}_{1})$. Finally, we set the complex phase of the CDW gap to zero.
\begin{figure}
\begin{centering}
\includegraphics[width=\columnwidth]{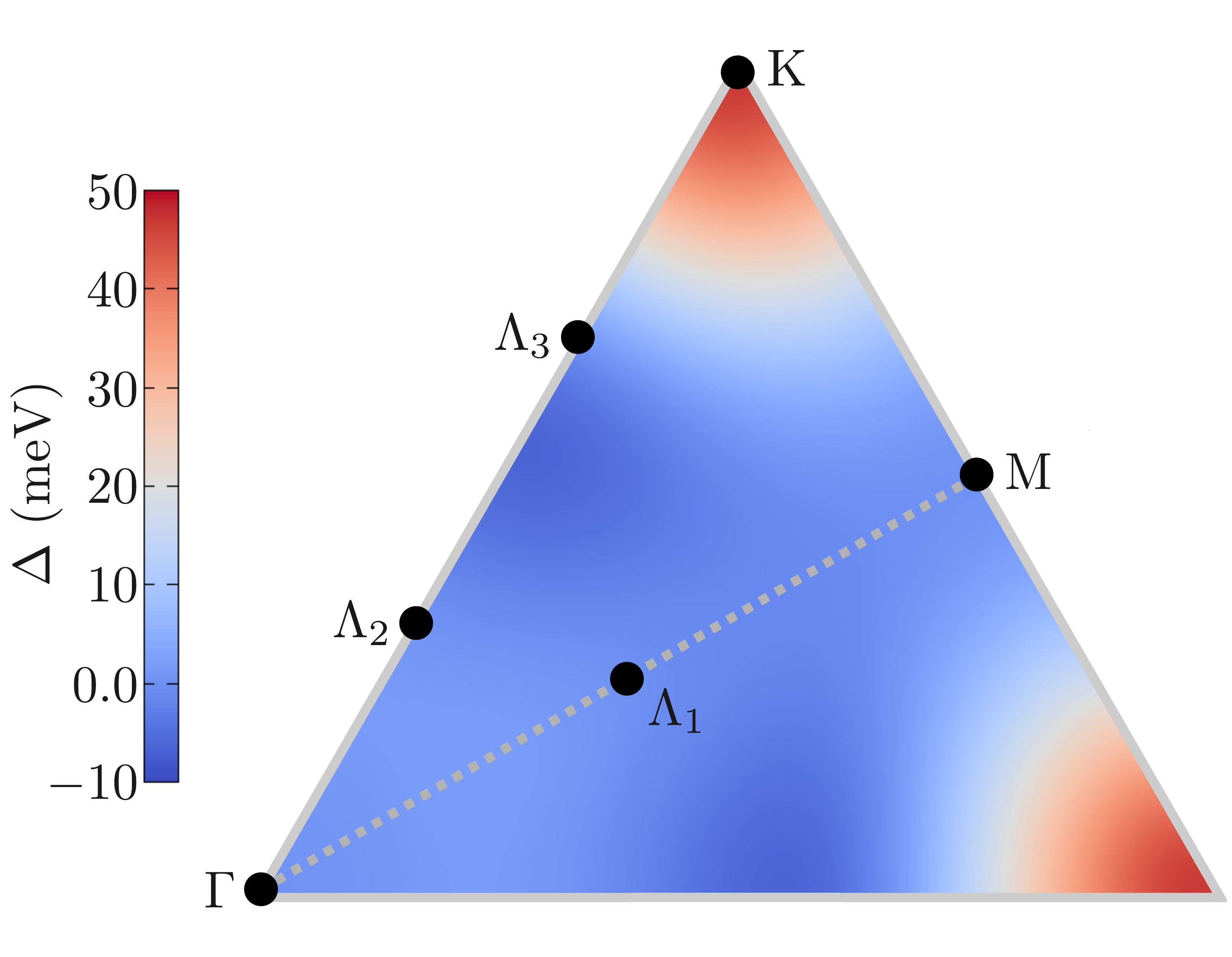}
\end{centering}
\caption{\label{fig:high sym}The self-consistent solution $\Delta(\mathbf{k})$ to the CDW gap equation, expanded to fifth order in periodic functions with the symmetry of the lattice. The points at which the function was evaluated in order to determine self-consistency of the solution are indicated.}
\end{figure}

The gap function can be expanded in terms of periodic functions with the same symmetry as the lattice~\cite{RahnEA12}, which up to fifth order results in the expression:
\begin{align}
\Delta_{\mathbf{k}} = & t_{0}+t_{1}\left(2\cos\left(\zeta\right)\cos\left(\eta\right)+\cos\left(2\zeta\right)\right)\notag \\
  & +t_{2}\left(2\cos\left(3\zeta\right)\cos\left(\eta\right)+\cos\left(2\eta\right)\right)\notag \\
  & +t_{3}\left(2\cos\left(2\zeta\right)\cos\left(2\eta\right)+\cos\left(4\zeta\right)\right)\notag \\
  & +t_{4}\left(\cos\left(\zeta\right)\cos\left(3\eta\right)+\cos\left(5\zeta\right)\cos\left(\eta\right) \right. \notag \\ 
& ~~~~~~~~~~ \left.+\cos\left(4\zeta\right)\cos\left(2\eta\right)\right) \notag \\
  & +t_{5}\left(2\cos\left(3\zeta\right)\cos\left(3\eta\right)+\cos\left(6\zeta\right)\right).\label{eq:tight_binding_D}
\end{align}
Here the definitions $\zeta = \frac{1}{2}k_{x}$ and $\eta = \frac{\sqrt{3}}{2}k_{y}$ have been used. The values of the six coefficients in this expansion are then calculated by searching for a self-consistent solution to the gap equation at six high-symmetry points in the first Brillouin zone. These points, and the resulting momentum-dependent gap function, are displayed in Fig.~\ref{fig:high sym}. Of the tested high-symmetry points the gap function was found to have a non-zero value only at K, in agreement with the observation in ARPES experiments that a CDW gap opens only in the Fermi sheets closest to the K-points~\cite{BorisenkoEA09}.

Having found the momentum-dependent gap function, we can examine the effect of the CDW formation in NbSe$_2$ on its electronic band structure and density of states. The former quantity is accessed experimentally in ARPES measurements, where the measured intensity is proportional to the electronic spectral function $A(\mathbf{k},\omega)$, which within the Lehmann representation is related to the renormalized electronic Green's function in the presence of a gap by the expression~\cite{Diagrammatics,Mahan}:
\begin{align}
G(\mathbf{k},i\omega_{n}) &=-\frac{1}{\pi}\int\mbox{d}\epsilon\frac{\mathfrak{Im}\left[G(\mathbf{k},\epsilon+i\delta)\right]}{i\omega_{n}-\epsilon} \notag \\
&=\int\mbox{d}\epsilon\frac{A(\mathbf{k},\epsilon)}{i\omega_{n}-\epsilon}.
\label{eq:Lehmann}
\end{align}
\begin{figure}
\begin{centering}
\includegraphics[width=\columnwidth]{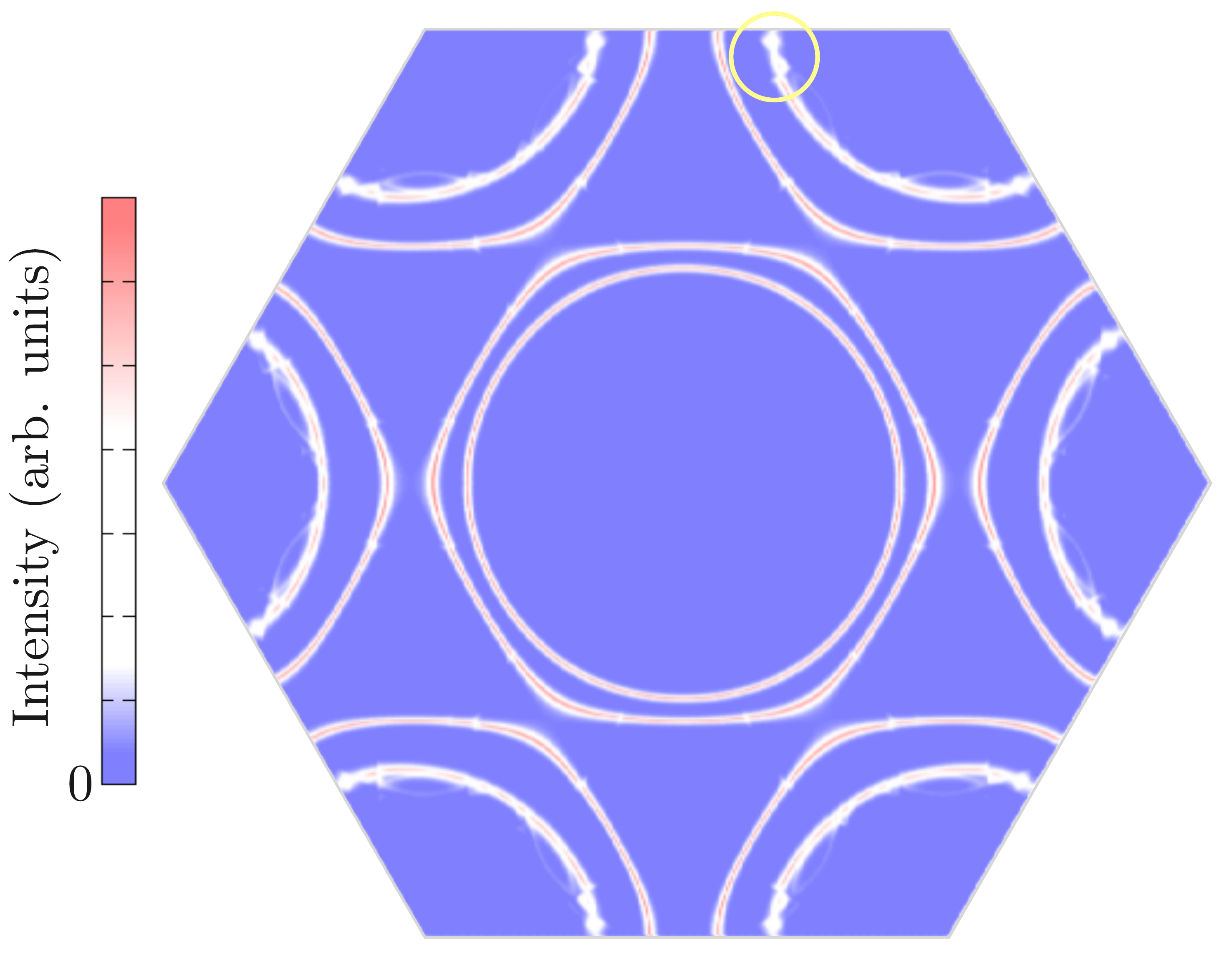}
\end{centering}
\caption{\label{fig:exptsFS}The calculated ARPES intensity at $E_{\text{F}}$ in the presence of the self-consistent CDW gap, plotted throughout the Brillouin zone. An electronic gap can be seen to suppress the intensity on regions of the inner band around the K-pockets, as highlighted by the yellow circle.
}
\end{figure}
\begin{figure*}
\begin{centering}
\includegraphics[width=\textwidth]{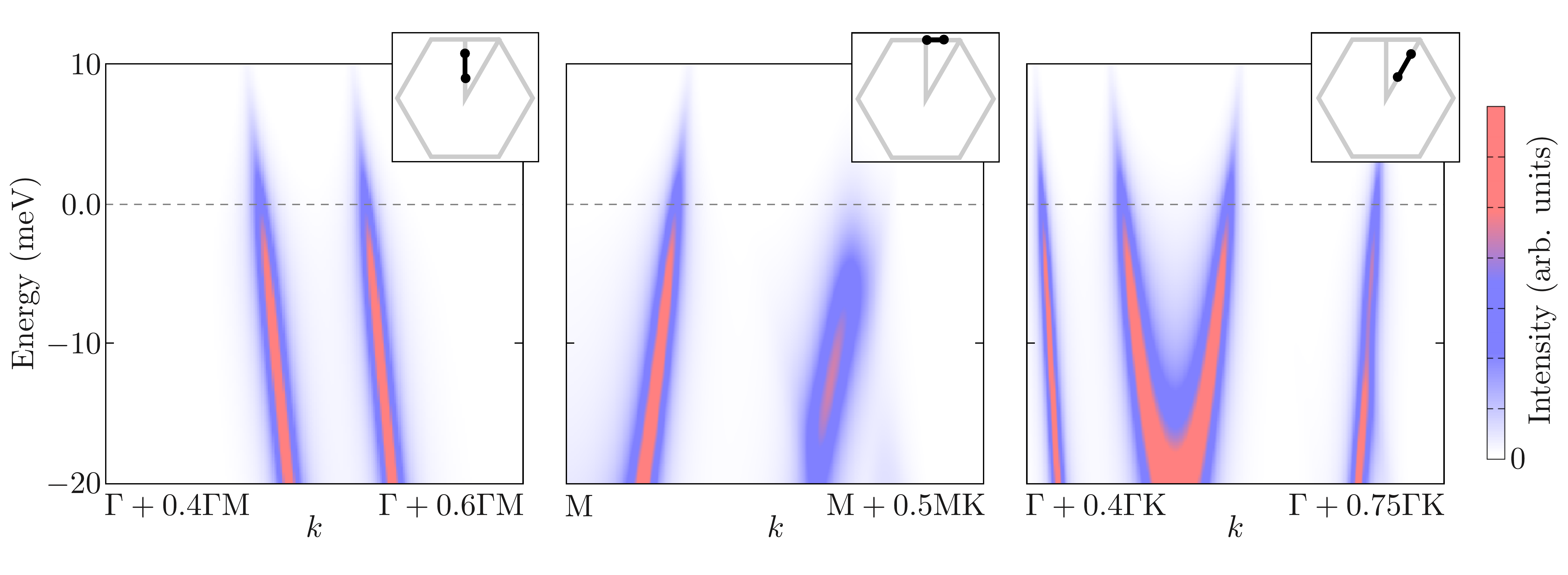}
\end{centering}
\caption{\label{fig:exptsDispersion}The calculated ARPES intensity in the presence of the self-consistent CDW gap along various high-symmetry lines. The CDW gap can be seen to open in the band closest to K along the MK line.}
\end{figure*}

Figure~\ref{fig:exptsFS} shows the calculated ARPES intensity at the Fermi energy as a function of momentum throughout the Brillouin zone. The outer pockets around the K-points do not develop a significant gap, while on the inner pockets a gap opens up along the MK lines. This is seen more clearly in Fig.~\ref{fig:exptsDispersion}, which shows the calculated intensity as a function of both energy and momentum along several high-symmetry lines. The opening of the gap on the inner pocket can again be clearly distinguished along the MK line. Notice that to make these figures, the spectral function was multiplied by a Fermi-Dirac distribution function, reflecting the fact that ARPES can only measure occupied electronic states. Both representations of the spectral function can be compared directly to experimentally-obtained ARPES intensity maps~\cite{BorisenkoEA09}. To obtain good quantitative as well as qualitative agreement between the calculation and experimental observations, a constant self energy of $7\,$meV was included in these plots.

\begin{figure}
\begin{centering}
\includegraphics[width=\columnwidth]{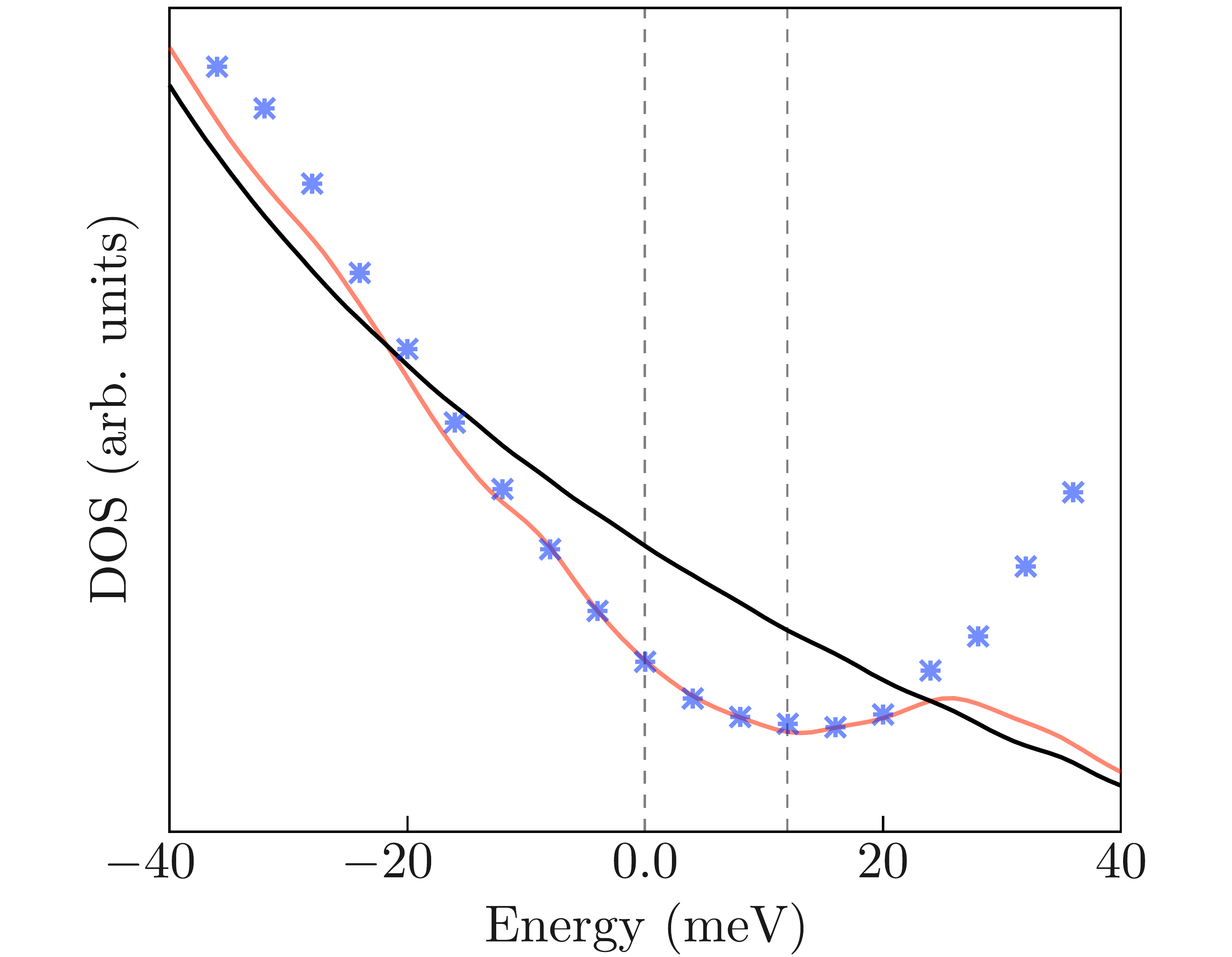}
\end{centering}
\caption{\label{fig:expts}Comparison of the modeled density of states to experimental results. Blue points: $dI/dV$ measured in STS experiments~\cite{SoumyanarayananEA13}. Black: the density of states calculated from the tight-binding fit to the electronic structure, without a CDW gap. Red: the calculated density of states including the self-consistent CDW gap.
}
\end{figure}

Figure~\ref{fig:expts} shows the computed density of states and compares it to that obtained from scanning tunneling spectroscopy (STS) measurements, where the derivative of the measured current with respect to the applied bias voltage is proportional to the local density of states~\cite{SoumyanarayananEA13}. The theoretical prediction matches the experimental observation close to the Fermi energy. In particular, both suggest a particle-hole asymmetric CDW gap centered around $12\,\mbox{meV}$ above $E_{\text{F}}$~\cite{SoumyanarayananEA13,FFJvWNatComms15}. To obtain a quantitative fit with the experimental data, we employed a $4\,\mbox{meV}$ shift of the chemical potential, which is well within the $\pm16\,\mbox{meV}$ uncertainty in the band structure fit~\cite{RahnEA12}. The kinks in the experimental results around $\pm35\,\mbox{meV}$ are due to the inelastic tunneling of electrons at these energies, exciting a phonon mode in the NbSe$_2$ surface~\cite{SoumyanarayananEA13}. 

Notice that the only free parameter in the model is the overall strength of the electron-phonon coupling. Its value was optimized to simultaneously match both the experimental ARPES and STS results. That good fits to all experimental data can be produced simultaneously by a single choice of the free parameter, as shown in figures~\ref{fig:exptsFS}-\ref{fig:expts}, indicates the significant contribution of the orbital- and momentum-dependent electron-phonon coupling in determining the momentum-resolved CDW gap, and hence the physical characteristics of the CDW state.

\section{Uniaxial strain\label{sec:Higher-Order-Diagrams} }
It was recently shown using scanning tunneling microscopy (STM) that regions of 1Q charge order appear on the surface of NbSe$_2$ alongside the usual 3Q CDW phase~\cite{SoumyanarayananEA13}. It was suggested that the 1Q regions may be stabilized by local strain on the sample's surface. The fact that both 1Q and 3Q order can be observed side-by-side on the same sample suggests that the 3Q order is in fact close to a quantum critical point, and can be destroyed by relatively little strain~\cite{SoumyanarayananEA13,FFJvWPRB15}. To quantify this hypothesis, we now compare the free energy of the 3Q ordered state to that of the single Q pattern, as a function of externally-applied uniaxial strain.

\begin{figure}
\begin{centering}
\includegraphics[width=\columnwidth]{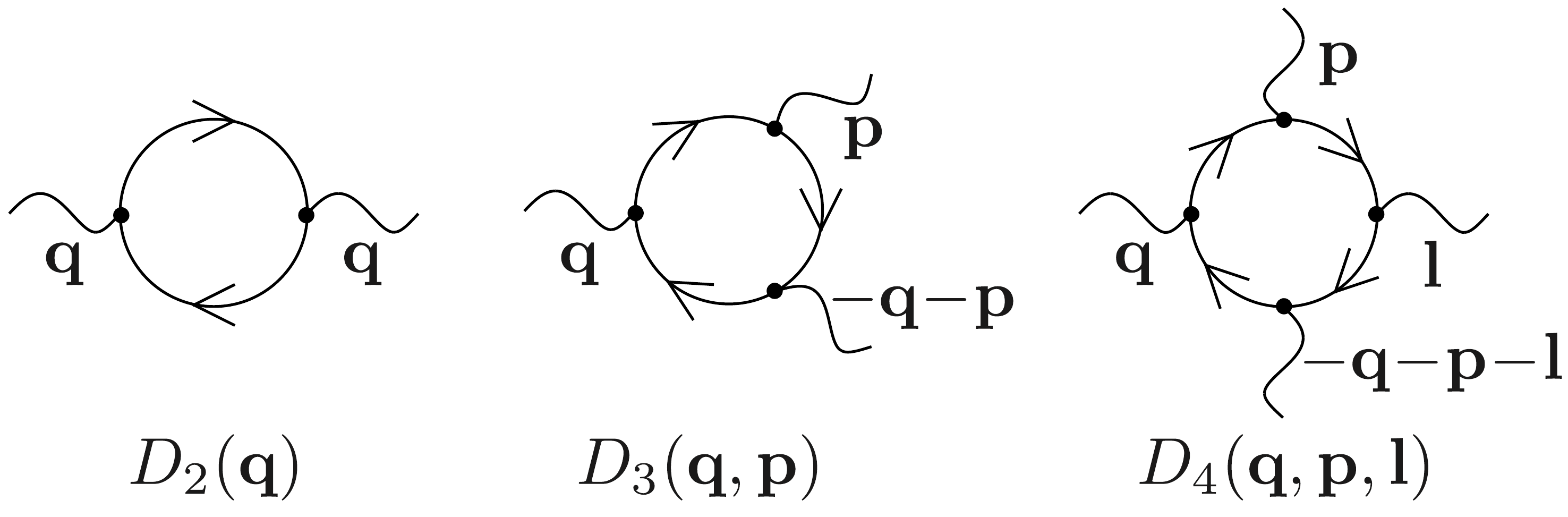}
\end{centering}
\caption{\label{fig:Feynman diagrams weak}The Feynman diagrams for the coefficients in the expansion of Eq.~\eqref{eq:Seff4}.}
\end{figure}

The free energy $F$ is related to the effective action through the relation:
\begin{align}
\exp\left(-\beta F\right) & = \int\mathscr{D}\varphi\exp\left(-S_{\text{eff}}\left[\varphi\right]\right).\label{eq:SeffF}
\end{align}
The effective action in this expression follows from the general expression of Eq.~\eqref{eq:Sfull} by integrating out the electronic degrees of freedom. It can expanded in powers of the electron-phonon coupling, with the expansion coefficients shown diagrammatically in Fig.~\ref{fig:Feynman diagrams weak}. The result up to fourth order is given in the static limit by:
\begin{align}
S_{\text{eff}}\left[\varphi\right] =&\sum_{q}\varphi_{q}^{\dagger}\left(\Omega_{\mathbf{q}}+\frac{1}{2}D_{2}\left(\mathbf{q}\right) \right)\varphi_{q}^{\phantom{\dagger}} \notag \\
& +\frac{1}{3}\sum_{\mathbf{qp}}\varphi_{\mathbf{q}}\varphi_{\mathbf{p}}\varphi_{-\mathbf{p}-\mathbf{q}}D_{3}\left(\mathbf{q},\mathbf{p}\right) \notag \\
& +\frac{1}{4}\sum_{\mathbf{pql}}\varphi_{\mathbf{q}}\varphi_{\mathbf{p}}\varphi_{\mathbf{l}}\varphi_{-\mathbf{l}-\mathbf{p}-\mathbf{q}}D_{4}\left(\mathbf{q},\mathbf{p},\mathbf{l}\right)
\label{eq:Seff4}
\end{align}
where we introduced the static nonlinear susceptibilities:

\begin{widetext}
\begin{align}
D_{3}\left(\mathbf{q},\mathbf{p}\right) &= \sum_{\mathbf{k},\mu,\nu,\rho}g_{\mathbf{k},\mathbf{k}+\mathbf{q}}^{\mu, \nu}g_{\mathbf{k}+\mathbf{q},\mathbf{k}+\mathbf{q}+\mathbf{p}}^{\nu,\rho}g_{\mathbf{k}+\mathbf{q}+\mathbf{p},\mathbf{k}}^{\rho,\mu} \left[\frac{f\left(\xi_{\mathbf{k}}^{\mu}\right)}{\left(\xi_{\mathbf{k}}^{\mu}-\xi_{\mathbf{k}+\mathbf{q}}^{\nu}\right)\left(\xi_{\mathbf{k}}^{\mu}-\xi_{\mathbf{k}+\mathbf{p}+\mathbf{q}}^{\rho}\right)}+\text{cyclic permutations}\right] \notag \\
D_{4}\left(\mathbf{q},\mathbf{p},\mathbf{l}\right) &= \sum_{\mathbf{k},\mu,\nu,\rho,\sigma}g_{\mathbf{k},\mathbf{k}+\mathbf{q}}^{\mu, \nu}g_{\mathbf{k}+\mathbf{q},\mathbf{k}+\mathbf{q}+\mathbf{p}}^{\nu,\rho}g_{\mathbf{k}+\mathbf{q}+\mathbf{p},\mathbf{k}+\mathbf{q}+\mathbf{p}+\mathbf{l}}^{\rho,\sigma}g_{\mathbf{k}+\mathbf{q}+\mathbf{p}+\mathbf{l},\mathbf{k}}^{\sigma,\mu} \cdot \notag \\ & ~~~~~~~~~~~~ \left[\frac{f\left(\xi_{\mathbf{k}}^{\mu}\right)}{\left(\xi_{\mathbf{k}}^{\mu}-\xi_{\mathbf{k}+\mathbf{q}}^{\nu}\right)\left(\xi_{\mathbf{k}}^{\mu}-\xi_{\mathbf{k}+\mathbf{q}+\mathbf{p}}^{\rho}\right)\left(\xi_{\mathbf{k}}^{\mu}-\xi_{\mathbf{k}+\mathbf{q}+\mathbf{p}+\mathbf{l}}^{\sigma}\right)}+\text{cyclic permutations}\right].
\label{eq:D3D4}
\end{align}
\end{widetext}
The apparent divergences in these expressions are canceled by the cyclic permutation over band indices, and can be removed analytically~\cite{MelikyanNorman14SuppMat}.

\begin{figure}
\begin{centering}
\includegraphics[width=\columnwidth]{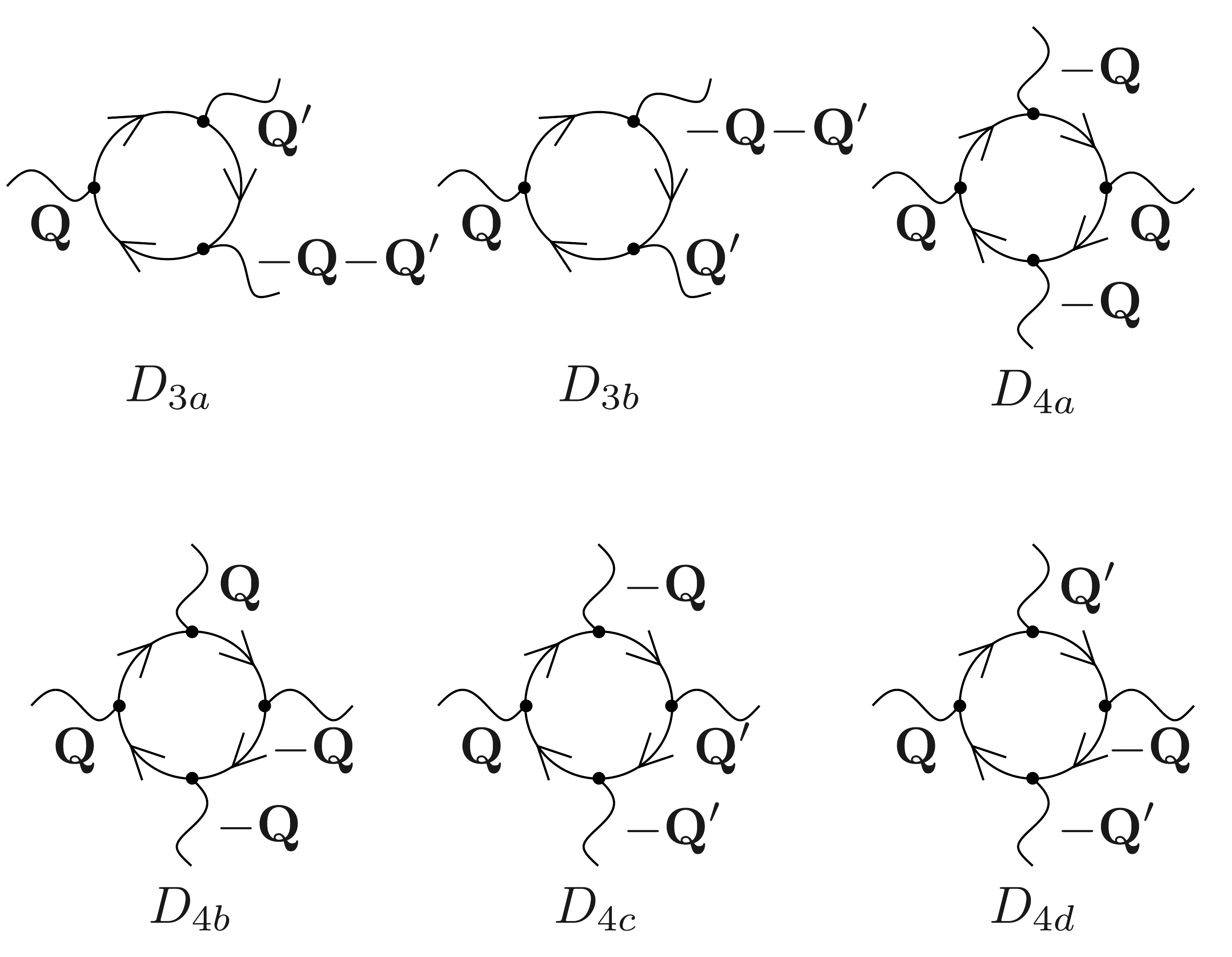}
\end{centering}
\caption{\label{fig:symmetryFactors}The Feynman diagrams remaining in the free energy expansion after attention has been restricted to the CDW vectors $\mathbf{Q}_{i}$. The labels $\mathbf{Q}$ and $\mathbf{Q}'$ indicate different CDW wave vectors.}
\end{figure}

To investigate the relative stability of the different ordering geometries that have been experimentally observed in NbSe$_2$, the general displacement fields $\varphi_{\mathbf{q}}$ can be replaced by variables which are nonzero only for momentum values $\mathbf{q}=\pm\mathbf{Q}_j$, with $\mathbf{Q}_j$ all possible CDW wave vectors. Writing $\varphi_{i} = \varphi\left(\mathbf{Q}_{i}\right)$ and neglecting the functional integral in Eq.~\eqref{eq:SeffF} then implies:
\begin{align}
\beta F = & S_{\text{eff}} = -r\sum_{i}\left|\varphi_{i}\right|^{2}-a\varphi_{1}\varphi_{2}\varphi_{3} \notag \\
&+\frac{1}{2}b\left(\sum_{i}\left|\varphi_{i}\right|^{4}+ c\sum_{i\neq j} \left|\varphi_{i}\right|^{2}\left|\varphi_{j}\right|^{2}\right).\label{eq:F(a,b,c)}
\end{align}
Notice that in this expression, the free energy has implicitly been minimized with respect to the relative phases of the displacement fields, resulting in the appearance of a minus sign in the cubic term. The expansion coefficients can be written explicitly in terms of the phonon energy, the electronic susceptibility, and the diagrams shown in Fig.~\ref{fig:symmetryFactors} as:
\begin{align}
r & = -\Omega_{\mathbf{q}}+D_{2}\notag \\
a & = 2\left(D_{3a}+D_{3b}\right)\notag \\
b & = D_{4a}+2D_{4b}\notag \\
bc & = 4\left(D_{4d}+2D_{4c}\right).\label{eq:rabc}
\end{align}

The free energy can be further simplified by focussing explicitly on the two experimentally-observed geometries of 3Q and 1Q order. The 3Q state is described by setting $\varphi_{j}=\varphi$ for all $j$, while the 1Q state is characterized by only a single non-zero displacement field, $\varphi_{1}=\varphi$. The expressions for the free energies then reduce to:
\begin{align}
\beta F_{\text{3Q}}\left[\varphi\right] & = -3r\varphi^{2}-a\varphi^{3}+\frac{3}{2}b\left(c+1\right)\varphi^{4} \notag \\
\beta F_{\text{1Q}}\left[\varphi\right] & = -r\varphi^{2}+\frac{1}{2}b\varphi^{4}.
\label{eq:F_3Q_1Q}
\end{align}
In all cases considered in this article, both $b$ and $b(c+1)$ are positive, and no higher order terms in the expansion need to be considered.

Minimizing the functionals of Eq.~\eqref{eq:F_3Q_1Q} with respect to $\varphi$ gives the free energy of each particular configuration. In the case of the 1Q CDW the result is:
\begin{align}
\beta F_{\text{1Q}}^{\text{min}}=\begin{cases} \begin{array}{c} -r^{2}/2b,\\ 0, \end{array} & \begin{array}{c} r>0\\ r\le0 \end{array}\end{cases}.
\label{eq:Fmin1}
\end{align}
The temperature dependence of the free energy is dominated by that of the coefficient $r$ in front of the quadratic term. It is then clear from the expression above that 1Q order may develop as a function of temperature through a second-order phase transition at $r=0$ when approaching from the disordered regime with $r<0$. 

For the 3Q case the situation is altered by the cubic term. The free energy of the ordered system then becomes:
\begin{align}
\beta F_{\text{3Q}}^{\text{min}} = -&\left[a^{4}+24ra^{2}b\left(c+1\right)+96r^{2}b^{2}\left(c+1\right)^{2} \right. \notag \\
 &~~ \left.+a\left(a^{2}+16rb\left(c+1\right)\right)^{\frac{3}{2}}\mbox{sgn}\left(ac+a\right)\right] / \notag \\
& \left[64b^{3}\left(c+1\right)^{3}\right].
\label{eq:Fmin3}
\end{align}
Again assuming that only the temperature dependence of $r$ is relevant near the transition, naively we would expect two possible transitions: either the 3Q order develops through a second-order phase transition at $r=0$, or it emerges from a first order phase transition when $b\left(c+1\right) = -a^{2}/18r$. In fact the second-order transition is always intercepted by a `weakly first-order' transition~\cite{LarkinPikin}. To see this, consider the disordered state with $r<0$. Upon cooling, $r$ rises towards zero. However, just before the temperature is low enough for $r$ to vanish and the second-order phase transition to be realized, it must be the case that $r\rightarrow0^{-}$. In this limit, for any nonzero $a$, the term $a^{2}/r$ diverges and passes through the value $-18b(c+1)$, at which point a first-order transition sets in. The second-order 3Q transition can therefore never occur.

For completeness, one should also consider the possibility of a 2Q CDW state rather than the 1Q and 3Q phases considered so far. The free energy corresponding to that case would be:
\begin{align}
\beta F_{\text{2Q}}\left[\varphi\right] = & -2r\varphi^{2}+\frac{b}{2}\left(c+2\right)\varphi^{4}.
\end{align}
Comparing this expression to the 1Q case, it immediately follows that $F_{\text{1Q}}^{\text{min}}=\left(c+2\right)F_{\text{2Q}}^{\text{min}}$. The 1Q geometry is thus energetically favorable to the 2Q geometry for any $c>-1$, and a 2Q phase never forms.

\begin{figure}
\begin{centering}
\includegraphics[width=\columnwidth]{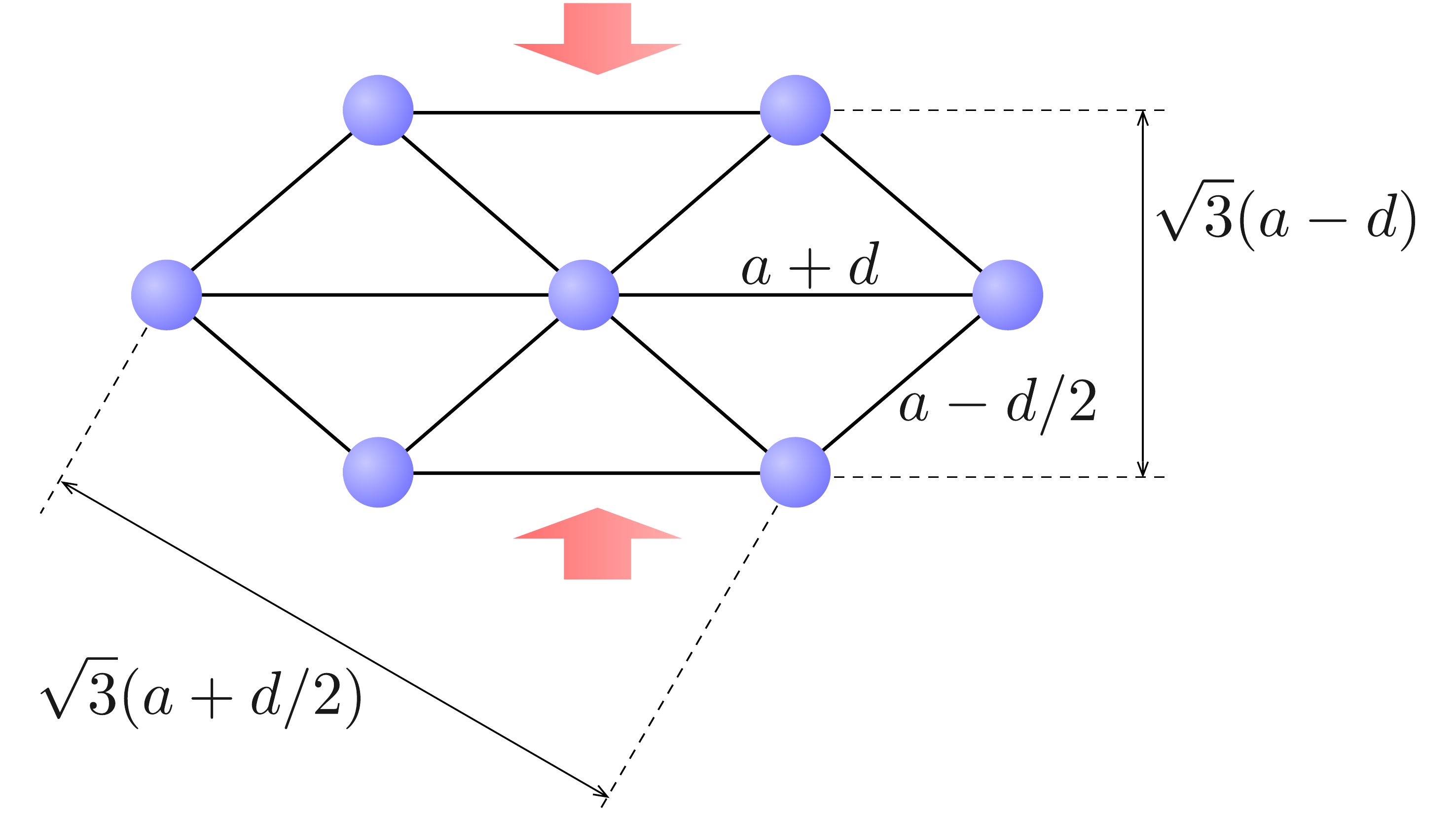}
\end{centering}
\caption{\label{fig:strain}The effect of uniaxial strain on lattice distances. Compression of the atomic lattice along one of the CDW propagation vectors leads to an expansion along the directions of the remaining CDW propagation vectors. The distances are indicated to first order in $d$.}
\end{figure}

Including the dependence on uniaxial strain, as well as temperature, in the free energies of Eqs.~\eqref{eq:Fmin1} and~\eqref{eq:Fmin3}, breaks the sixfold symmetry of the disordered and $3Q$ states. One way in which the asymmetry manifests itself is through a difference in the phonon frequencies along stretched and compressed directions within the crystal. Assuming the volume of the unit cell to be conserved, an expansion of the lattice in one direction will be accompanied by a corresponding compression in the orthogonal direction. The uniform-volume assumption corresponds to a Poisson's ratio of unity; this is unlikely to be realistic, but may still be used as a reasonable first approximation. Further assuming the phonon energies to depend linearly on changes in the atomic spacing, at the small displacements considered here, the strain dependence can be included in the free energy functionals by making the substitutions $\Omega_{1}\rightarrow\Omega_{1}\left(1+\sigma\right)$ and  $\Omega_{2,3}\rightarrow\Omega_{2,3}\left(1-\frac{\sigma}{2}\right)$. Here $\sigma$ is a dimensionless strain parameter, and the relation between the displacements along different CDW propagation directions is shown graphically in Fig.~\ref{fig:strain}.

The effect of strain on the quadratic coefficients in the free energy expansion arises from the different effect it has on the phonon dispersion in different directions, which leads to the inequivalent terms:
\begin{align}
r_{1} &= -\Omega_{\mathbf{q}}\left(1+\sigma\right)+D_{2} \notag \\
r_{2,3} &= -\Omega_{\mathbf{q}}\left(1-\frac{\sigma}{2}\right)+D_{2}.
\label{eq:r123}
\end{align}
Since the diagrams in Fig.~\ref{fig:Feynman diagrams weak} contain no dependence on internal phonon lines, they do not depend on the phonon frequency $\Omega_{\mathbf{q}}$. Besides the quadratic coefficients, therefore, no other terms in the free energy are affected by the applied uniaxial strain.
\begin{figure}
\begin{centering}
\includegraphics[width=\columnwidth]{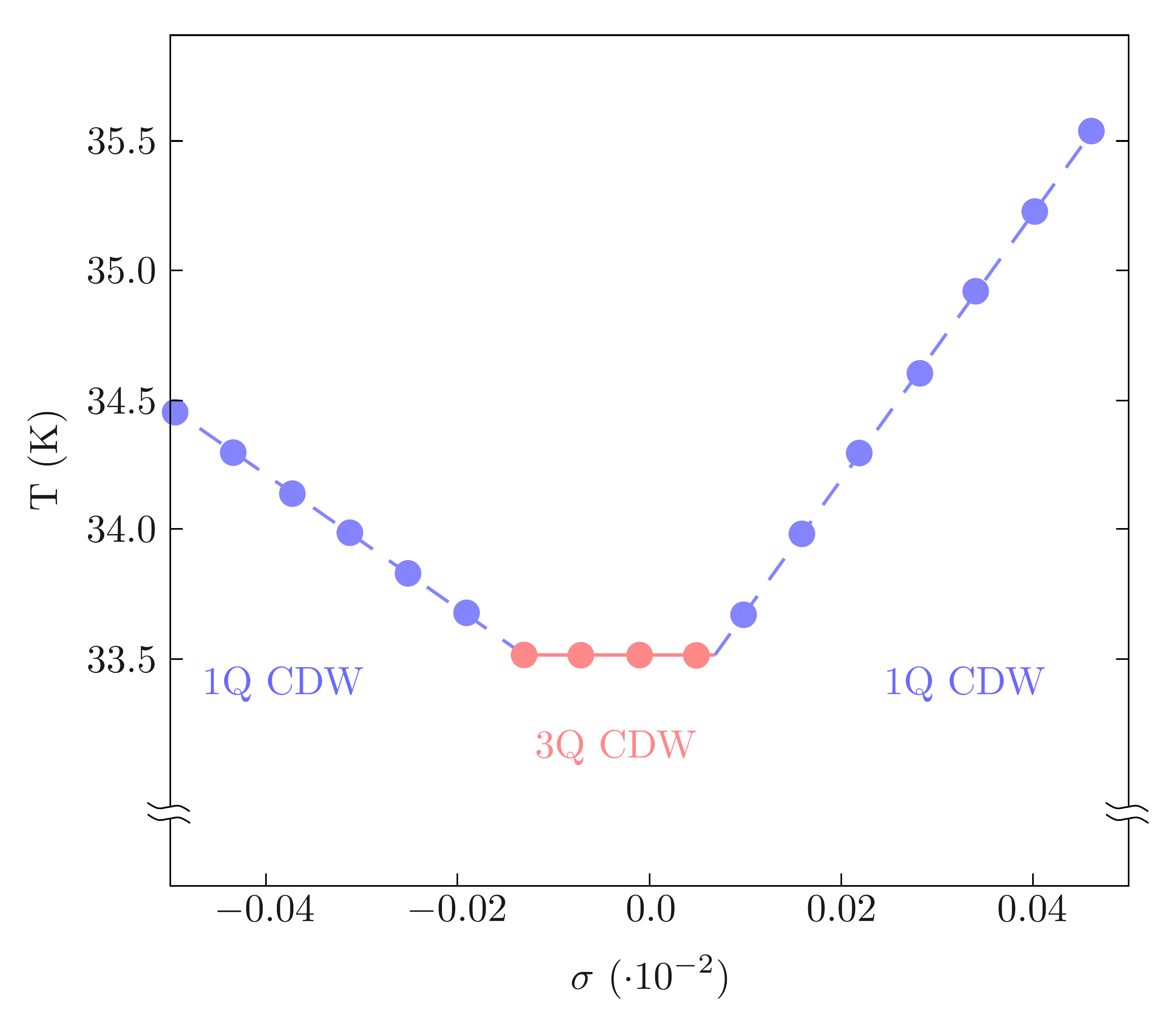}
\end{centering}
\caption{\label{fig:RPA_phase_diagram}The phase diagram arising from the mean-field free energy expansion, as a function of temperature and strain. Dashed lines bounding the 1Q regions indicate second-order phase transitions whereas the solid line bounding the 3Q region indicates a weakly-first-order transition.}
\end{figure}

Combining both the temperature and strain dependence of the free energy leads to the phase diagram displayed in Fig.~\ref{fig:RPA_phase_diagram}. To obtain this diagram, the magnitude of the electron-phonon coupling $g$ has been chosen such that the zero-strain phase transition occurs at the experimentally-observed transition temperature $T_{\text{CDW}}=33.5\,$K. The propagation vectors for the 3Q and 1Q phases are set equal to the experimentally-established values $\mathbf{Q} = 0.657\, \Gamma$M and $\mathbf{Q} = 2/7 \,\Gamma$M respectively~\cite{SoumyanarayananEA13}. The different rates of increase of the transition temperature with increasing positive or negative strain in Fig.~\ref{fig:RPA_phase_diagram} are due to the fact that a compression (positive strain) $\sigma$ along the $\mathbf{Q}_1$ direction causes only a $\sigma/2$ stretch (negative strain) in the directions of the other two possible CDW wave vectors. 

While the 3Q geometry is stable at zero strain, it only takes around $0.01\%$ anisotropic change in the bare phonon energy to break the symmetry down to 1Q. An estimate for the corresponding lattice strain would require knowledge of the in-plane Gr\"{u}neisen parameter,$-\partial\ln\Omega_{\mathbf{q}}/\partial\ln a$, with in-plane lattice parameter $a$. In the layered hexagonal materials MoS$_{2}$, BN, and graphite, this parameter is known to be of order unity throughout most of the first Brillouin zone~\cite{ConleyEA13,MounetEA05}. Assuming the trend holds for NbSe$_{2}$, the phase diagram of Fig.~\ref{fig:RPA_phase_diagram} indicates that around $0.01\%$ uniaxial strain of the atomic lattice suffices to stabilize the 1Q geometry. This may be compared to the upper bound of around $0.45\%$ lattice strain estimated for the 1Q-ordered, locally-strained regions in the STM experiment~\cite{SoumyanarayananEA13}. The quantitative discrepancy will shortly be shown to be due to the neglect of lattice fluctuations in the mean-field approximation used up to this point.

The 3Q CDW phase in NbSe$_2$ is thus found to be intrinsically close to a quantum phase transition into a 1Q-ordered state. This result suggests that besides occurring in spontaneously-formed locally-strained regions on the surface of samples used in STM studies~\cite{SoumyanarayananEA13}, the 1Q CDW may also be stabilized in bulk crystals under the application of only a moderate uniaxial strain. The quantum critical region separating the two ordering geometries should be experimentally accessible in the same way.

\section{The CDW Pseudogap\label{sec:pseudogap}}
Including the momentum and orbital dependence of the electron-phonon coupling in a mean-field theory based on a tight-binding fit to the electronic band structure, as used up to this point, has elucidated various ground state properties of the CDW order in NbSe$_2$. It cannot, however, be expected to give a full description of the high-temperature physics. It is well-known that in the presence of strong electron-phonon coupling, the entropy associated with fluctuating atomic displacements significantly contributes to the properties of CDW materials at finite temperatures~\cite{VarmaSimons83}. In NbSe$_2$ such fluctuations may be expected to be present, because of the experimentally-established pseudogap-like properties of the disordered state just above the CDW transition temperature. In close analogy to the pseudogap state of high-temperature superconductors, these include the formation of Fermi arcs as seen in ARPES experiments~\cite{BorisenkoEA09}, and the stabilization of locally-ordered regions surrounding defects in the bulk disordered state seen in STM experiments~\cite{ArguelloEA14}.

\begin{figure}[h]
\begin{centering}
\includegraphics[width=\columnwidth]{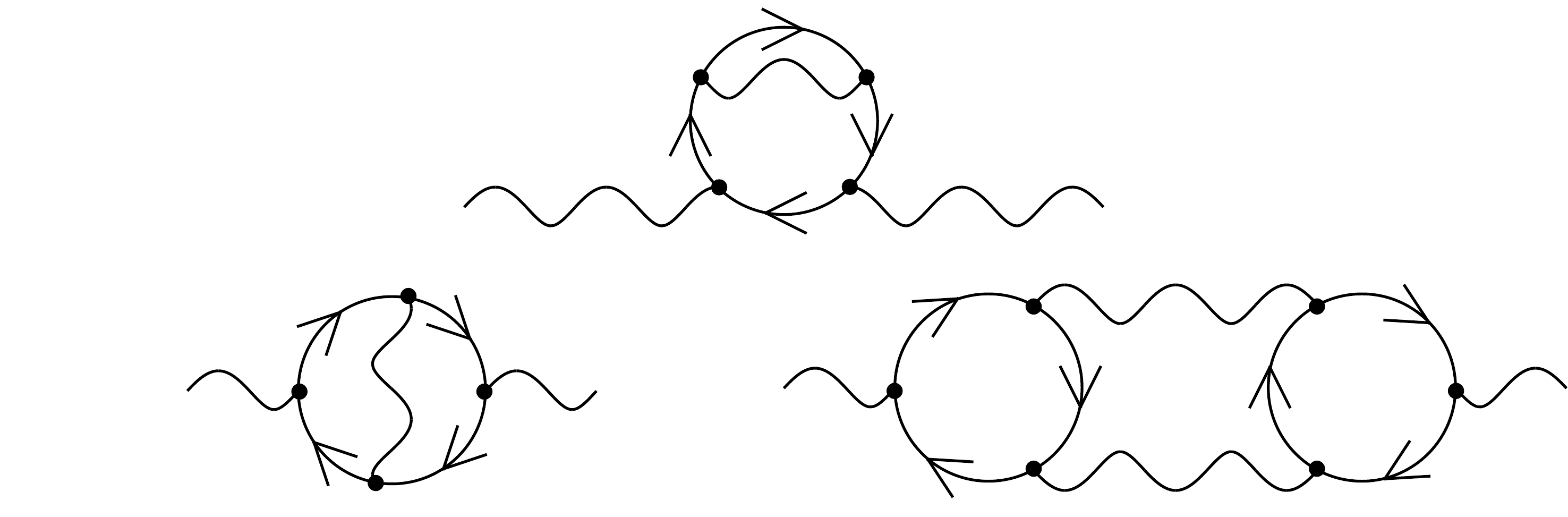}
\end{centering}
\caption{\label{fig:MMA}The Feynman diagrams beyond those of Fig.~\ref{fig:Feynman diagrams weak}, which contribute to the free energy expansion within the mode-mode coupling approximation (MMA). The top diagram provides the dominant contribution in the case of NbSe$_2$.}
\end{figure}

Fluctuations in the phonon field are neglected in the RPA approximation employed up to this point, and to include them we need to go beyond the mean field diagrams of Fig.~\ref{fig:Feynman diagrams weak} in the evaluation of the free energy coefficients. The lowest-order contributions including fluctuations of the phonon field are shown diagrammatically in Fig.~\ref{fig:MMA}. Taking into account only these additional terms in the free energy expansion is known as the mode-mode coupling approximation (MMA)~\cite{Inglesfield80,VarmaSimons83,YoshiyamaEA86}.

The bottom-right diagram in Fig.~\ref{fig:MMA} is of order $g^{6}$, while the other contributions are of order $g^{4}$. In some cases, such as that of spin density waves in two-dimensional metals, the presence of Landau damping can suppress a factor of $g^{2}$ in this term and put it on equal footing with the other two~\cite{ChubukovEA14}. In the present case of CDW order, however, it is simply a sixth-order term in the electron-phonon coupling, and may be neglected~\cite{VarmaSimons83,YoshiyamaEA86}.

In the bottom-left diagram of Fig.~\ref{fig:MMA} the internal phonon line constitutes a vertex correction. Such corrections may be dismissed for vertices with large momentum transfer by appeal to Migdal's theorem, which guarantees they will be proportional to an additional ratio of electronic to ionic mass~\cite{Migdal58,Schrieffer}. For the case of NbSe$_2$, the momentum transfer of interest is that along the CDW wave vector with $\left|\mathbf{Q}_{\text{CDW}}\right|\approx 2/3 \Gamma$M, and this diagram may therefore be neglected as well.

The remaining, top, diagram in Fig.~\ref{fig:MMA} can be thought of as a renormalization of the phonon propagator, but in contrast to the mean-field expression it includes a self-energy for one of the particles in the virtual electron-hole pair. As noted before, this self energy can be calculated separately using the expression depicted diagrammatically in Fig.~\ref{fig:Feynman}. The analytic form of the expression is given in the Appendix. In the self-energy diagrams, the phonon lines may be interpreted to represent the RPA-renormalized phonon propagators found before, while the renormalized electron lines can be found self-consistently by the inclusion of the self energy:
\begin{align}
D(\mathbf{q},i\Omega_{n}) &=\frac{-2\Omega_{\mathbf{q}}}{\left(i\Omega_{n}\right)^{2}-\Omega_{\mathbf{q}}^{2}+\Omega_{\mathbf{q}}D_{2}(\mathbf{q},i\Omega)} \notag \\
G(\mathbf{k},i\omega_{n}) &=\left(i\omega_{n}-\xi_{\mathbf{k}}-\Sigma_{k}+\mu\right)^{-1}.\label{eq:G self-energy}
\end{align}
Owing to the independence of the two bands crossing the Fermi level, only intra-band contributions need to be considered, and band indices are dropped for clarity.

Using the definitions of Eq.~\eqref{eq:G self-energy}, the expressions for the electronic self-energy $\Sigma$ and the generalized susceptibility $D_{2}^{\Sigma}$ in the presence of the fluctuations described by the MMA terms can be cast into a convenient form using the Lehmann representation of Eq.~\eqref{eq:Lehmann}:
\begin{widetext}
\begin{align}
\Sigma(\mathbf{k},\epsilon+i\delta) &=  \sum_{\mathbf{q}}\left|g_{\mathbf{k},\mathbf{k}-\mathbf{q}}\right|^{2}\left(\frac{\Omega_{\mathbf{q}}}{\Omega_{\text{RPA}}}\right)\int\mbox{d}\epsilon' A(\mathbf{k}-\mathbf{q},\epsilon')\left\{\frac{n_{B}(\Omega_{\text{RPA}})+1-f(\epsilon')}{\epsilon-\epsilon'-\Omega_{\text{RPA}}+i\delta}+\frac{n_{B}(\Omega_{\text{RPA}})+f(\epsilon')}{\epsilon-\epsilon'+\Omega_{\text{RPA}}+i\delta}\right\} \notag \\
D_{2}^{\Sigma}(\mathbf{q},i\Omega) &= \sum_{\mathbf{k}}\left|g_{\mathbf{k},\mathbf{k}-\mathbf{q}}\right|^{2}\int\mbox{d}\epsilon A(\mathbf{k},\epsilon)\mathfrak{Re}\left[\frac{f(\epsilon)-f(\xi_{\mathbf{k}-\mathbf{q}})}{\epsilon-\xi_{\mathbf{k}-\mathbf{q}}-i\Omega}\right].
\label{eq:D2(Sigma)}
\end{align}
\end{widetext}
In this expression, $\Omega_{\text{RPA}}$ are the RPA-renormalized phonon energies, which can be found by considering the poles of $D(\mathbf{q},i\Omega_{n})$. A self-consistent value for the electronic self energy can be found by solving the system of Eqs.~\eqref{eq:G self-energy}-\eqref{eq:D2(Sigma)}. Starting from the ansatz of a purely-real $\Sigma^{\left(0\right)}=7\,$meV, a full iteration of the self-consistent equations yields only a small correction to the self energy, suggesting this ansatz provides a reasonable approximation of the true self-consistent solution.

\begin{figure}
\begin{centering}
\includegraphics[width=\columnwidth]{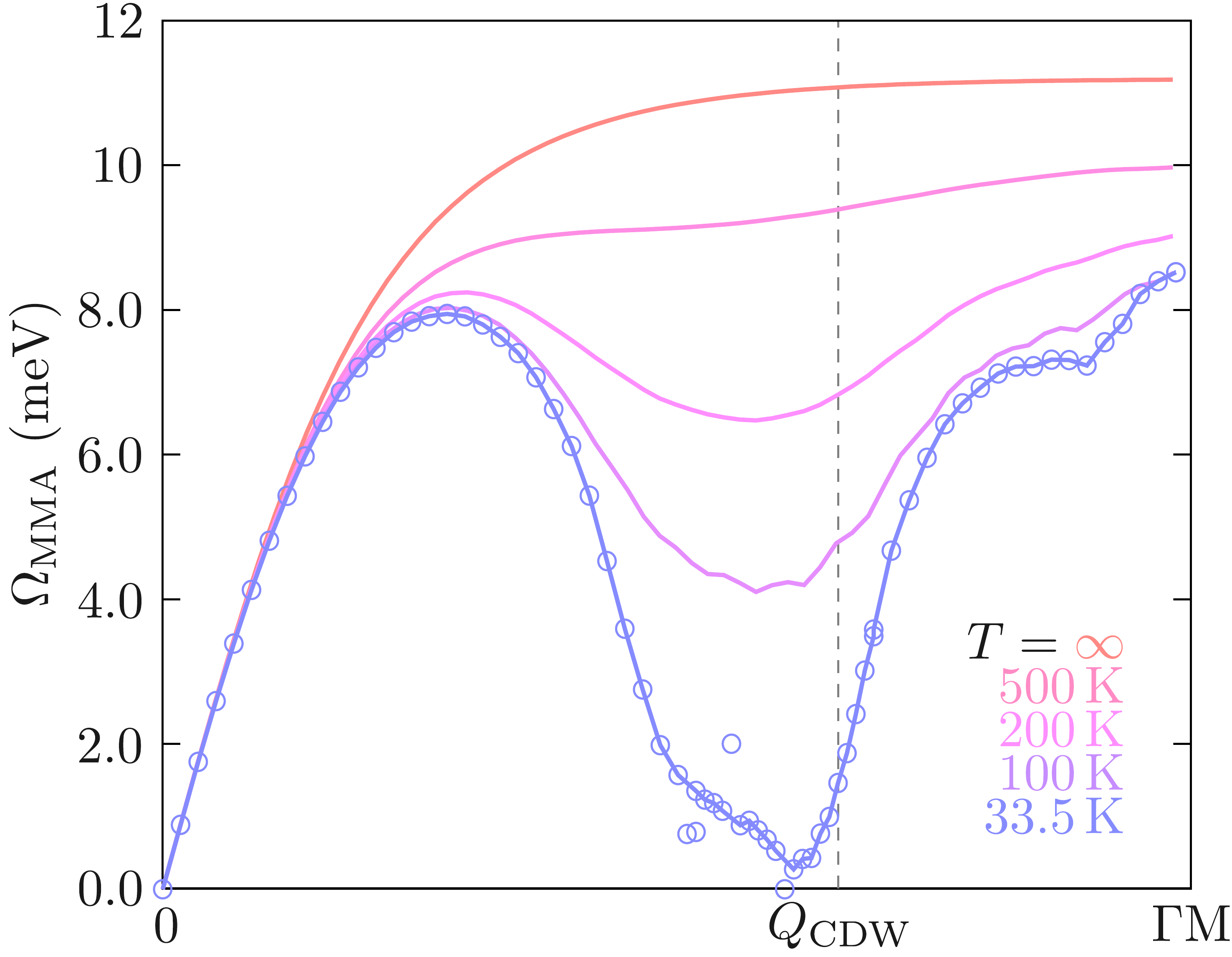}
\end{centering}
\caption{\label{fig:O_MMA O_RPA}The renormalized phonon dispersion along the high-symmetry direction $\Gamma$M as calculated within the mode-mode coupling (MMA) approximation, at various temperatures. The $33.5\,$K data were calculated with a higher density $k$-space mesh. In all cases, the magnitude of the electron-phonon coupling has been set to give the MMA phase transition at $T=33.5\,$K.}
\end{figure}

Taking into account the phonon fluctuations and their associated self energy results in the renormalized phonon dispersion shown in Fig.~\ref{fig:O_MMA O_RPA}. The overall strength of the electron-phonon coupling $g$ has been chosen such that the MMA-renormalized dispersion touches zero energy at the experimentally-observed CDW transition temperature $T_{\text{CDW}}=33.5\,\mbox{K}$. Calculating the RPA-renormalized phonon dispersion using the same value for $g$ would result in a far more heavily renormalized (lower energy) phonon frequency. The RPA in fact predicts an ordering transition at $416\,\mbox{K}$ with this value of $g$. The effect of the fluctuations encoded by the additional MMA diagrams is therefore to suppress the CDW order.

Physically, the suppression of the CDW transition to lower temperatures can be interpreted in terms of phase fluctuations~\cite{LeeEA73,McMillan77}. At the mean field transition temperature $T_{\text{RPA}} = 416\,\mbox{K}$, the CDW order parameter acquires a non-zero amplitude, signaled by the finite expectation value $\sqrt{\langle|\varphi|^{2}\rangle }>0$. The displacement field has a phase as well as an amplitude, however, and fluctuations of the phase cause the order parameter itself to average to zero: $\left\langle \varphi\right\rangle =0$. In the calculation of the phonon dispersion, these fluctuations and their detrimental effect on the CDW order are taken into account by the MMA diagrams. Only at the much lower temperature $T_{\text{MMA}} =33.5\,\mbox{K}$ are the fluctuations sufficiently damped to allow for long-range order with $\left\langle \varphi\right\rangle >0$. 

This result can be extended to include the effects of uniaxial strain in the same way that strain was introduced at the mean field level. Within the mode-mode coupling approximation, only the coefficient of the quadratic term in the free energy obtains additional contributions beyond RPA. As in Eq.~\eqref{eq:r123}, that coefficient can be written as a combination of the energy $\Omega_{\mathbf{q}}$ required to excite a phonon in a given direction, and the electronic susceptibility $D_2$. In this case, however, the MMA contribution to the electronic susceptibility shown in the middle of Fig.~\ref{fig:MMA} includes an internal phonon line, and thus becomes direction-dependent as well. It can be found by first considering the anisotropic electronic self energy, which is obtained from Eq.~\eqref{eq:D2(Sigma)} after replacing the phonon energy by either one of the strained versions $\Omega_{\mathbf{q}}(1+\sigma)$ or $\Omega_{\mathbf{q}}(1-\sigma/2)$. Inserting the anisotropic self energy into the expression for $D_2$ in Eq.~\eqref{eq:D2(Sigma)} then results in an anisotropic version of the electronic susceptibility, which finally can be substituted into the coefficient for the quadratic term in the free energy.

\begin{figure}
\begin{centering}
\includegraphics[width=\columnwidth]{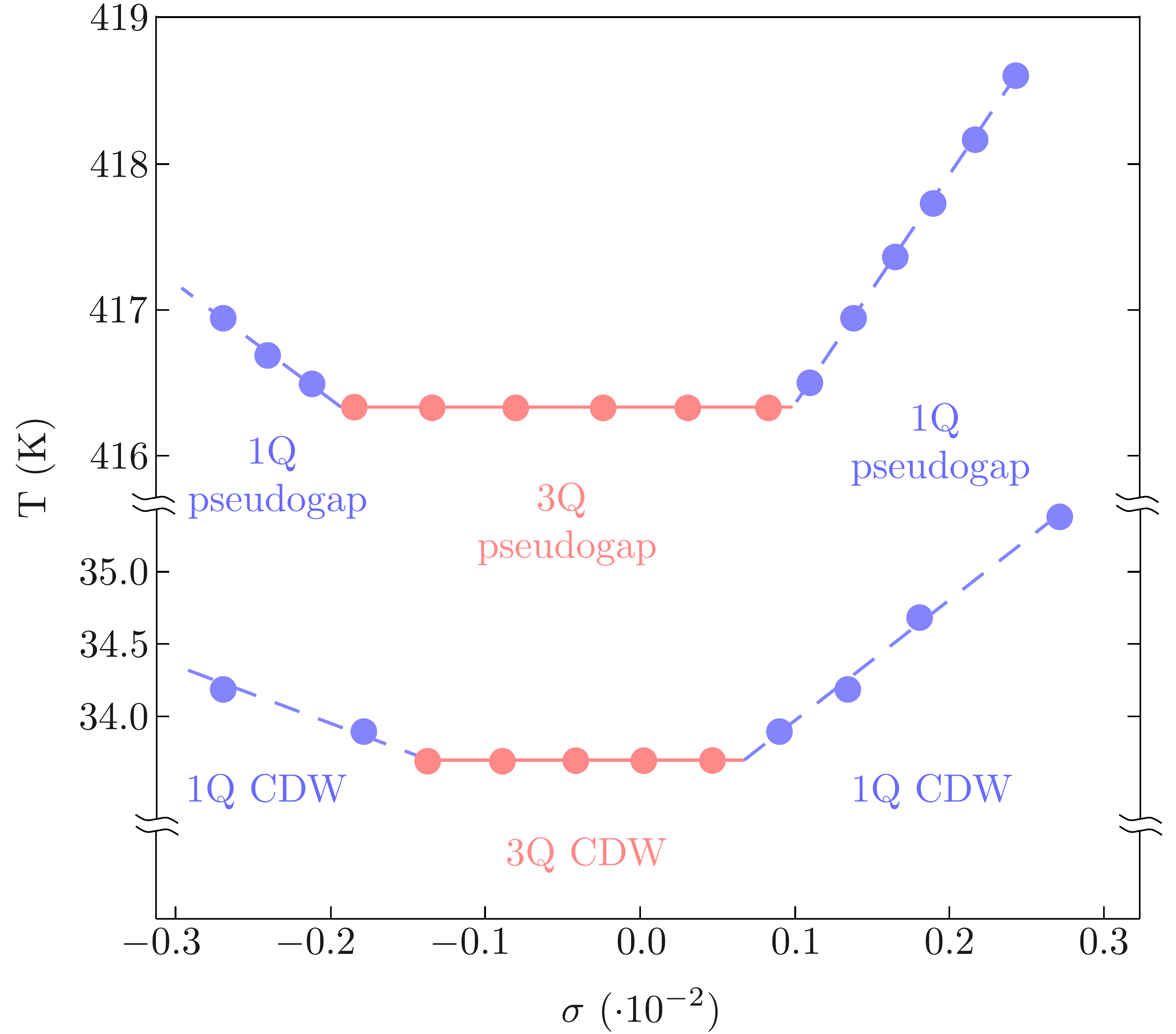}
\end{centering}
\caption{\label{fig:MMA phase diagram} The calculated phase diagram as a function of temperature and uniaxial strain. The pseudogap regime indicates the temperatures below the point at which the mean-field (RPA) renormalised phonon dispersion first touches zero energy, but above the CDW phase transition as calculated within the mode-mode coupling (MMA) approximation. Dashed lines indicate second-order phase transitions whereas solid lines indicate weakly-first-order transitions within the respective approximations. The onset of the pseudogap phase is expected to be a crossover rather than a true phase transition.}
\end{figure}

Combining the result of the minimization procedure outlined above, with that using the mean-field approximation, results in the temperature-strain phase diagram shown in Fig.~\ref{fig:MMA phase diagram}. The overall strength of the electron-phonon coupling $g$ is chosen so as to give a phase transition temperature $T_{\text{MMA}}$ equal to the experimentally-observed value of $T_{\text{CDW}}=33.5\,$K. The figure also shows the mean-field transition temperatures $T_{\text{RPA}}$ found using the same value of the electron-phonon coupling. The RPA and MMA transition temperatures were calculated independently, rather than within a single framework. That is, within the random phase approximation, the transition temperature is predicted to be $T_{\text{RPA}}$, and the free energy expansion breaks down below that temperature. Within the mode-mode coupling approximation, nothing happens at $T_{\text{RPA}}$, and long-range order sets in at the much lower temperature $T_{\text{MMA}}$. Physically, we expect the region between these two temperatures to constitute a pseudogap regime, in which local fluctuations of the phase of the order parameter, present in the MMA, prevent the emergence of long-range order, even though the amplitude of the order parameter which is present already within RPA has a non-zero expectation value. 

Notice that the presence of fluctuations has the additional effect of stabilizing the 3Q ordered phase at low temperatures up to lattice strains of around $0.1\%$, as opposed to $0.01\%$ in the mean-field calculation of figure~\ref{fig:RPA_phase_diagram}. The MMA value is consistent with the upper bound of about $0.45\%$ lattice strain estimated for the 1Q-ordered, locally-strained regions in the STM experiment~\cite{SoumyanarayananEA13}. 

The picture of charge order in NbSe$_2$ arising out of a locally-fluctuating state is consistent with several recent experimental observations~\cite{ChatterjeeEA14}. First of all, in the pseudogap regime above $T_{\text{CDW}}$, the closed loops forming the 2D Fermi surface can be seen to break up into arcs~\cite{BorisenkoEA09}. These gaps develop from the true CDW gap below the transition temperature, and seem to persist to temperatures well above the experimental limit of $119\,\mbox{K}$~\cite{BorisenkoEA09}. The presence of this pseudogap in the absence of any long-range CDW order indicates that the CDW amplitude, which is directly proportional to the gap size, retains a non-zero value at high temperatures even though the order parameter itself has disappeared.

Secondly, scanning tunneling experiments have shown that static, local CDW order surrounds defects on the surface of NbSe$_2$ up to temperatures of at least $96\,\mbox{K}$~\cite{ArguelloEA14}. The defects in these images are interpreted as stabilizing CDW order locally, and suppressing temporal fluctuations. They do so, however, in an uncorrelated manner, maintaining the overall absence of long-range order. Following the minima of the renormalized phonon dispersions in Fig.~\ref{fig:O_MMA O_RPA} as a function of temperature, gives an indication of the expected thermal evolution of the CDW vector ${Q}_{\text{CDW}}$ characterizing the local order in the fluctuating islands of the pseudogap phase~\cite{FFJvWPRB15}. The regions of stability surrounding defects are expected to grow continuously as temperature decreases, until at $T_{\text{CDW}}$ they overlap sufficiently for separate regions to become correlated. Besides the islands of static order stabilized by defects and seen by STM, there must also exist dynamically-fluctuating short-ranged order which is invisible on the long timescales required by STM. It may be assumed that the correlation length characterizing the dynamical fluctuations is comparable to that of the static islands. This is corroborated by the observation within X-ray diffraction of a finite but non-zero correlation length for the CDW order at high temperatures, which becomes truly long-ranged only at $T_{\text{CDW}}$~\cite{ChatterjeeEA14}.

The CDW pseudogap in NbSe$_2$ is thus interpreted as a regime with non-zero amplitude of the CDW order parameter and its associated electronic gap, but with long-range phase coherence suppressed by local phase fluctuations. The fluctuations originate in the atomic displacement field, and may be suppressed around defects, leading to observable islands of static order well above the CDW transition temperature. The amplitude of the CDW order parameter itself, and hence also the electronic gap and local order, will disappear at the mean field transition temperature, which, based on the phase diagram of Fig.~\ref{fig:MMA phase diagram}, is expected to be of order $400\,\mbox{K}$.

\section{Discussion\label{sec:Conclusions}}
In this paper we presented in detail a model for the charge density wave order in $2H$-NbSe$_2$, extending previous work investigating the gap and pseudogap regimes~\cite{FFJvWNatComms15} and the stability of the 1Q and 3Q geometries under uniaxial strain~\cite{FFJvWPRB15}. Employing a strong electron-phonon coupling which depends both on the ingoing and outgoing electron momenta, and on the orbital character of the bands scattered between, we found quantitative agreement between the model predictions and a number of experimental observations. 

This model accounts for the presence of a CDW gap in only one of the two bands crossing the Fermi level by relative strength of its intraband electron-phonon coupling as compared to all other couplings. We found that the generalized electronic susceptibility, which includes the effect of the full electron-phonon coupling, is peaked at the experimentally-observed CDW wave vector, removing the seemingly-mysterious situation of an incommensurate CDW arising from a practically-flat bare electronic susceptibility. We quantified the contribution of electronic nesting to the selection of the CDW wave vector by calculating the value of its nesting parameter, ranging from zero (no nesting) to one (perfect quasi-1D nesting), finding the value $\alpha=0.55$. This value indicates that for any realistic description of the CDW phase in NbSe$_2$, both the electronic structure and the momentum and orbital dependent electron-phonon coupling are indispensable ingredients.

By explicitly calculating the momentum dependence of the CDW gap, taking into account the shapes of the electronic dispersion and the electron-phonon coupling, we found quantitative agreement with experimental observations of the electronic density of states around $E_{\text{F}}$~\cite{SoumyanarayananEA13}, as well as the gapped electronic bands seen in ARPES experiments~\cite{BorisenkoEA09}. The fact that fixing only a single fitting parameter leads to simultaneous agreement with all available data connected to the CDW gap structure emphasizes once more the importance of the interplay between electronic structure and electron-phonon coupling~\cite{FFJvWNatComms15}. From the density of states, it can be seen that the CDW gap in NbSe$_2$ is offset by about $12\,$meV from $E_{\text{F}}$, which explains the difference in reported gap sizes throughout the literature~\cite{SoumyanarayananEA13}. The distribution of the gap in momentum space indicates that it is concentrated mainly in the region where the inner pocket surrounding the K-point crosses the MK-line. This feature explains why the CDW phase appears to be associated in ARPES experiments with Fermi arcs~\cite{BorisenkoEA09}. The fact that these arcs persist in the experiment even at high temperatures can be explained by the inclusion of fluctuations of the atomic displacements. Such fluctuations tend to suppress CDW order, and a pseudogap regime is found between the predicted mean-field transition temperature and the actual CDW transition in the presence of fluctuations. The pseudogap regime is characterized by having a non-zero value for the order parameter amplitude, and hence the gap size, but without long-range phase coherence. This state is characterized by Fermi arcs in the electronic structure, and short-ranged, locally-fluctuating order in the charge density, in agreement with all available relevant experimental observations~\cite{BorisenkoEA09,ArguelloEA14,ChatterjeeEA14}.

Finally, we included the effect of externally-applied uniaxial strain by means of an anisotropic contribution to the phonon energy~\cite{FFJvWPRB15}. The resulting phase diagram indicates that NbSe$_2$ is naturally close to a quantum phase transition between the observed 3Q CDW order at zero strain, and a unidirectional 1Q phase at about $0.1\%$ applied strain. In addition to the spontaneously-formed regions of locally-strained 1Q CDW order which have been observed on the surface of NbSe$_2$ samples in STM experiments~\cite{SoumyanarayananEA13}, this result suggests that the transition into unidirectional order, and the corresponding quantum critical region, may be accessible in experiments employing bulk uniaxial strain.

Both the similarity of the locally-fluctuating high-temperature phase to the pseudogap phase observed in cuprate high-T$_C$ superconductors, and the natural vicinity of the material to a quantum critical transition between different ordering geometries, make NbSe$_2$ an ideal model system for the study of strongly-correlated, charge-ordered materials. The fact that all of the experimentally-accessible properties in both the charge-ordered and pseudogap regimes can be understood quantitatively within a single one-parameter model based on a strong, momentum- and orbital-dependent, electron-phonon coupling, opens the way for comparison to related material families.

\section*{Acknowledgments}
FF~acknowledges support from a Lindemann Trust Fellowship of the English Speaking Union. JvW~acknowledges support from a VIDI grant financed by the Netherlands Organisation for Scientific Research (NWO).

%%%Appendix in wide text:
\begin{widetext}

\section*{Appendix: Electronic self energy}
The expression for the electronic self-energy displayed diagrammatically in Fig.~\ref{fig:Feynman} can be written as:
\begin{align}
\Sigma\left(\mathbf{k},i\omega_{n}\right)=-\sum_{\mathbf{q}}\sum_{\Omega_{n}}\left|g_{\mathbf{k},\mathbf{k}-\mathbf{q}}\right|^{2}G\left(\mathbf{k}-\mathbf{q},i\omega_{n}-i\Omega_{n}\right)D\left(\mathbf{q},i\Omega_{n}\right).
\end{align}
Here $D$ the is RPA-renormalized phonon propagator and $G$ is the fully-renormalized electronic propagator which itself depends on the self-energy $\Sigma$. This expression can be expanded and written in terms of the spectral function as:
\begin{align}
\Sigma\left(\mathbf{k},i\omega_{n}\right) & = -\sum_{\mathbf{q}}\sum_{i\Omega_{n}}\left(\left|g_{\mathbf{k},\mathbf{k}-\mathbf{q}}\right|^{2}\frac{-2\Omega_{0}\left(\mathbf{q}\right)}{\left(i\Omega_{n}+\Omega_{\text{RPA}}\left(\mathbf{q}\right)\right)\left(i\Omega_{n}-\Omega_{\text{RPA}}\left(\mathbf{q}\right)\right)} \int\mbox{d}\epsilon'\frac{A\left(\mathbf{k}-\mathbf{q},\epsilon'\right)}{i\omega_{n}-i\Omega_{n}-\epsilon'}\right).
\end{align}
The summation over Matsubara frequencies can then be carried out to yield:
\begin{align}
\Sigma\left(\mathbf{k},i\omega_{n}\right) & = \sum_{\mathbf{q}}\left|g_{\mathbf{k},\mathbf{k}-\mathbf{q}}\right|^{2}\left(\frac{\Omega_{0}}{\Omega_{\text{RPA}}}\right)\int\mbox{d}\epsilon' A\left(\mathbf{k}-\mathbf{q},\epsilon'\right)\left\{ \frac{n_{B}\left(\Omega_{\text{RPA}}\right)+1-f\left(\epsilon'\right)}{i\omega_{n}-\epsilon'-\Omega_{\text{RPA}}}+\frac{n_{B}\left(\Omega_{\text{RPA}}\right)+f\left(\epsilon'\right)}{i\omega_{n}-\epsilon'+\Omega_{\text{RPA}}}\right\}.
\end{align}
Wick rotating $i\omega_{n}\rightarrow\epsilon+i\delta$ and inserting the seed $\Sigma^{(0)}$ for the self energy, the result, $\Sigma^{(1)}$, after one iteration of the self-consistent calculation, is given by:
\begin{align}
\Sigma^{\left(1\right)}\left(\mathbf{k},\epsilon\right)  = & -\frac{1}{\pi}\sum_{\mathbf{q}}\int\mbox{d}\epsilon'\left|g_{\mathbf{k},\mathbf{k}-\mathbf{q}}\right|^{2}\frac{\Sigma^{\left(0\right)''}}{\left(\epsilon'-\xi_{\mathbf{k}-\mathbf{q}}-\Sigma^{\left(0\right)'}\right)^{2}+\left(\Sigma^{\left(0\right)''}-\delta\right)^{2}}\notag \\
&~~~~~~~~~~~~~~~\frac{\Omega_{0}}{\Omega_{\text{RPA}}} \left\{\frac{n_{B}\left(\Omega_{\text{RPA}}\right)+1-f\left(\epsilon'\right)}{\epsilon-\epsilon'-\Omega_{\text{RPA}}+i\delta}+\frac{n_{B}\left(\Omega_{\text{RPA}}\right)+f\left(\epsilon'\right)}{\epsilon-\epsilon'+\Omega_{\text{RPA}}+i\delta}\right\} .
\end{align}

\end{widetext} 
% end of Appendix.

\bibliographystyle{apsrev4-1}
\bibliography{mybib}

\end{document}